\documentclass[10pt]{article}
\usepackage{bm}
\usepackage{tikz}
\usetikzlibrary{decorations.pathreplacing,decorations.markings}
\usetikzlibrary{matrix}

\usepackage{amsthm,amsfonts,amsmath,amssymb}
\usepackage{pxfonts}
\usepackage{euscript}
\usepackage{hyperref}

\usepackage{xypic}
\usepackage{graphicx}
\usepackage{epstopdf} 
\usepackage{enumerate}
\usepackage{bbold}

\newcommand\void[1]       {}

\newtheorem{thm}{Theorem}

\newtheorem{prop-ph}[thm]{Proposition$^{\mathrm{ph}}$}
\newtheorem{cor-ph}[thm]{Corollary$^{\mathrm{ph}}$}
\newtheorem{lemma-ph}[thm]{Lemma$^{\mathrm{ph}}$}
\newtheorem{thm-ph}[thm]{Theorem}

\theoremstyle{definition}
\newtheorem{defn}[thm]{Definition}
\newtheorem{rem}[thm]{Remark}
\newtheorem{expl}[thm]{Example}

\newtheorem{defn-ph}[thm]{Definition$^{\mathrm{ph}}$}

\numberwithin{equation}{section}
\numberwithin{thm}{section}

\newcommand\nn             {\nonumber \\}

\newcommand\be            {\begin{equation}}
\newcommand\ee            {\end{equation}}
\newcommand\bea           {\begin{eqnarray}}
\newcommand\eea         {\end{eqnarray}}
\newcommand\bnu          {\begin{enumerate}}
\newcommand\enu          {\end{enumerate}}

\newcommand\id            {\mathrm{id}}

\newcommand\rep     {\EuScript{R}\mathrm{ep}}

\newcommand\one    {\mathbf{1}}

\newcommand{\Is}  {\mathbf{Ising}}
\newcommand{\toric} {\mathbf{Toric}}
\newcommand{\BH} {\mathbf{H}}
\newcommand{\Mod} {\mathrm{Mod}}

\newcommand{\pf}{\begin{proof}}
\newcommand{\epf}{\end{proof}}

\newcommand\Cb            {\mathbb{C}}

\newcommand\Zb            {\mathbb{Z}}

\newcommand\CA           {\EuScript{A}}
\newcommand\CB           {\EuScript{B}}
\newcommand\CC           {\EuScript{C}}
\newcommand\CD           {\EuScript{D}}

\newcommand\CM          {\EuScript{M}}

\newcommand\CX         {\EuScript{X}}

\def\Tr{\text{Tr}}

\makeatletter
\newcommand\xleftrightarrow[2][]{%
  \ext@arrow 9999{\longleftrightarrowfill@}{#1}{#2}}
\newcommand\longleftrightarrowfill@{%
  \arrowfill@\leftarrow\relbar\rightarrow}
\makeatother

\tikzset{middlearrow/.style={
        decoration={markings,
            mark= at position 0.5 with {\arrow{#1}} ,
        },
        postaction={decorate}
    }
}

\newcommand{\ket}[1]{|#1 \rangle}
\newcommand{\dd}{\mathrm{d}}
\newcommand{\ii}{\mathrm{i}}

\newcommand{\diag}[3][0pt]{\vcenter{\hbox{\raisebox{#1}{\includegraphics[height=#3]{#2}}}}}

\begin{document}

\begin{center} \LARGE
A topological phase transition on the edge of the 2d $\Zb_2$ topological order
\\
\end{center}

\vskip 2em
\begin{center}
{ Wei-Qiang Chen$^{a}$, 
Chao-Ming Jian$^{b}$,
Liang Kong$^{a}$,
Yi-Zhuang You$^{c}$, 
Hao Zheng$^{a,d}$
~\footnote{Emails:
{\tt  chenwq@sustech.edu.cn, cmjian@kitp.ucsb.edu, kongl@sustech.edu.cn, yzyou@ucsd.edu, hzheng@math.pku.edu.cn}}}
\\[1.8em]
$^a$ Shenzhen Institute for Quantum Science and Engineering, \\
and Department of Physics, \\
Southern University of Science and Technology, Shenzhen 518055, China
\\[0.8em]
$^b$ Kavli Institute for Theoretical Physics, University of California Santa Barbara, CA 93106, USA 
\\[0.8em]
$^c$ Department of Physics, University of California, San Diego, CA 92093, USA
\\[0.8em]
$^d$ Department of Mathematics, Peking University, Beijing, 100871, China
\end{center}

\vskip 2.5em

\begin{abstract}
The unified mathematical theory of gapped and gapless edges of 2d topological orders was developed by two of the authors. It provides a powerful tool to study pure edge topological phase transitions on the edges of 2d topological orders (without altering the bulks). In particular, it implies that the critical points are described by enriched fusion categories. In this work, we illustrate this idea in a concrete example: the 2d $\Zb_2$ topological order. In particular, we construct an enriched fusion category, which describes a gappable non-chiral gapless edge of the 2d $\Zb_2$ topological order; then use an explicit lattice model construction to realize the critical point and, at the same time, all the ingredients of this enriched fusion category.

\end{abstract}

\tableofcontents

\section{Introduction}

The subject of topological order has attracted a lot of attentions in recent years among physicists. The main reason is that topological orders are new phases of matter that go beyond Landau's paradigm of phases and phase transitions (see a recent review \cite{wen2} and references therein). Landau's paradigm is based on a symmetry-broken theory. The mathematical theory of symmetry is that of groups. The new phases of matter challenge us to find radically new mathematical language and tools to understand topological phases and phase transitions. In this work, we show that the critical point of a pure edge topological phase transition between two gapped edges of the same 2d topological order\footnote{We use ``$n$d'' to represent the spatial dimension and ``$n+$1D'' to represent the spacetime dimension.} can be precisely described by a mathematical structure called an enriched fusion category \cite{mp,kz2}. 

\medskip


A gapped edge of a 2d anomaly-free topological order can be viewed as an anomalous 1d topological order. It contains no local observables (such as correlation functions) in the long wave length limit except topological excitations, which can be fused among themselves and form a unitary fusion category (UFC) \cite{kitaev-kong,kong-anyon}. A topological phase transition restricted on the 1d edge (without altering the 2d bulk) will be called a pure edge topological phase transition (see for example \cite{pmn}). Observe that the gap in a neighborhood of the edge must be closed at the critical point. Therefore, 
\begin{quote}
the critical point of a pure edge topological phase transition should be nothing but a gappable non-chiral gapless edge of a 2d anomaly-free topological order. 
\end{quote}
As a consequence, to ask for a precise mathematical description of the critical points of pure edge topological phase transitions is equivalent to ask for that of the gappable non-chiral gapless edges of 2d anomaly-free topological orders.


\medskip
Based on a mixture of physical intuition and mathematical arguments, two of the authors established in a previous work \cite{kz2} a unified mathematical theory of both gapped and gapless edges of 2d anomaly-free topological orders. This theory allows us to treat gapped, chiral gapless and (gappable) non-chiral gapless edges on an equal footing. More precisely, it says that all physical observables on a gapped/gapless edge form an enriched fusion category. Therefore, the critical point of a pure edge topological phase transition, or a gappable non-chiral gapless edge, is precisely described by an enriched fusion category. A more complete mathematical theory of (gappable) non-chiral gapless edges will be given in \cite{kz4}.  

\medskip
In this work, we demonstrate this idea explicitly in a concrete example: a pure edge phase transition between the two different gapped edges of the 2d $\Zb_2$ topological order \cite{kitaev1,bk}. The main result of this work is summarized below. 
\bnu
\item In Section\,\ref{sec:edge}, we give an explicit construction of a gappable non-chiral gapless edge of the 2d $\Zb_2$ topological order given by the following triple
$$
(V\otimes_\Cb \overline{V}, \,\, \Is\boxtimes\overline{\Is}, \,\, (\Is\boxtimes\overline{\Is})_A),
$$
where 
\bnu
\item $V$ is the Ising chiral algebra or vertex operator algebra (VOA) of central charge $\frac{1}{2}$ and $\overline{V}$ is the same VOA but contains only anti-chiral fields $\phi(\bar{z}), \forall \phi\in V$; 

\item $\Is$ is the unitary modular tensor category (UMTC) of $V$-modules, i.e. $\Is=\Mod_V$. It contains three simple objects $\one, \psi, \sigma$ (with the fusion rule $\sigma\otimes \sigma = \one \oplus \psi$); $\overline{\Is}$ is the same tensor category as $\Is$ but with the braidings defined by the anti-braidings in $\Is$; 

\item $\Is\boxtimes\overline{\Is}$ is nothing but the Drinfeld center of $\Is$, i.e. 
$$
Z(\Is)=\Is\boxtimes\overline{\Is}.
$$

\item $A=\one\boxtimes \one \oplus \psi \boxtimes \psi$ is a condensable algebra in $Z(\Is)$ (see Definition\,\ref{def:separable}), and $Z(\Is)_A$ is the UFC of right $A$-modules in $Z(\Is)$. 

\enu
The enriched fusion category is determined by the pair $(Z(\Is), \,\, Z(\Is)_A)$ via the standard construction \cite{mp}. More precisely, 
\begin{itemize}
\item the objects in the enriched fusion category are the same as those in $Z(\Is)_A$;
\item the space of morphism $\hom(x,y)$ is defined by the internal hom $[x,y]=(y\otimes_A x^\ast)^\ast$ in $Z(\Is)$. The internal hom $[x,y]$ can be interpreted either as the domain wall between two boundary CFT's with boundary conditions $x$ and $y$, respectively, or as the partition function of a CFT defined on a strip with two boundary conditions $x$ and $y$ (see Figure\,\ref{fig:partition}). 
\end{itemize}

\item In Section\,\ref{sec:lattice}, we construct a lattice model to realize the critical point of the pure edge phase transition between two gapped edges of the $\Zb_2$ topological order, and, at the same time, all ingredients $[x,y]$ of the enriched fusion category.  
  
\enu
The significance of this work is two-fold: (1) it shows that the mathematical theory in \cite{kz2} can indeed provide a mathematical theory of all pure edge topological phase transitions; (2) it provides the first direct lattice model proof of the mathematical theory in \cite{kz2}. 

\medskip
Since the (enriched) categorical language is not so familiar to condensed matter physicists and the lattice model construction might not be so familiar to mathematical physicists, in order to invite readers from both communities, we have tried to be self-contained by recalling some basic things and a brief introduction of the theory in \cite{kz2}.

\bigskip
\noindent {\bf Acknowledgement}: WQC is supported by National Key Research and Development Program of China (No. 2016YFA0300300), and NSFC (No. 11674151 and No. 11861161001). CMJ is supported by the Gordon and Betty Moore Foundations EPiQS Initiative through Grant GBMF4304. LK and HZ are supported by the Science, Technology and Innovation Commission of Shenzhen Municipality (Grant Nos. ZDSYS20170303165926217 and JCYJ20170412152620376) and Guangdong Innovative and Entrepreneurial Research Team Program (Grant No. 2016ZT06D348). HZ is also supported by NSFC under Grant No. 11871078.

\section{Categorical Preliminaries}
In this section, we review some basic ingredients of a unitary modular tensor category (UMTC), and give two examples,  and set our notations along the way. 

\subsection{Unitary Modular Tensor Categories}
It is well-known that an anomaly-free 2d topological order without symmetry is described by a pair $(\CC, c)$, where $\CC$ is a UMTC of anyons and $c$ is the chiral central charge \cite{kitaev2}. 

\medskip
We review some important ingredients of a UMTC $\CC$. It has finitely many simple objects (simple anyons). We denoted the simple objects by $i, j, k\in \mathrm{Irr}(\CC)$, where $\mathrm{Irr}(\CC)$ is the finite set of the equivalence classes of simple objects. A generic object in $\CC$ is a direct sum of simple ones, e.g. $i\oplus j \oplus k$, and is called a composite anyon. For each pair $(x,y)$ of objects, the hom space $\hom_\CC(x,y)$ is a finite dimensional Hilbert space. It has a tensor product functor $\otimes: \CC \times \CC \to \CC$, i.e. $(x,y) \mapsto x\otimes y, \forall x,y\in \CC$, such that it is associative, i.e. there exists an isomorphism $x\otimes (y \otimes z) \xrightarrow{\alpha_{x,y,z}} (x\otimes y) \otimes z$ for all $x,y,z\in \CC$ satisfying necessary coherence conditions (i.e. pentagon relation). This data can be reduced to the induced linear isomorphisms $\hom_\CC(i\otimes (j\otimes k), l) \simeq \hom_\CC((i\otimes j) \otimes k, l)$ for $i,j,k,l\in \mathrm{Irr}(\CC)$, or equivalently, to $F$-matrices, which satisfies pentagon identities. The dimension of $\hom(i\otimes j, k)$, denoted by $N_{ij}^k$, is called the fusion rule. The category $\CC$ has a tensor unit $\one$, which is simple, together with unit isomorphisms $\one \otimes x \xrightarrow{l_x} x \xleftarrow{r_x} x\otimes \one$ for all $x\in \CC$ satisfying necessary coherence conditions (i.e. triangle relation). It has a unitary structure. More precisely, for each morphism $f: x\to y$, there is an adjoint morphism $f^\dagger: y\to x$ such that
\bea
&(g\otimes h)^\dagger=g^\dagger \otimes h^\dagger,\quad\quad \forall g: v\to w, h: x \to y, & \\
&\alpha_{x,y,z}^\dagger=\alpha_{x,y,z}^{-1},\quad l_x^\dagger=l_x^{-1},\quad r_x^\dagger=r_x^{-1}. & \label{eq:unitary-asso-unit}
\eea

For each $x\in \CC$, there is a dual object $x^\ast$ (the anti-particle of $x$), together with the duality morphisms (the creation/annihilation operators) expressed graphically as follows: 
\be
\begin{array}{llll}
  \raisebox{-8pt}{
  \begin{picture}(26,22)
   \put(0,6){\scalebox{.75}{\includegraphics{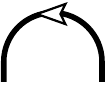}}}
   \put(0,6){
     \setlength{\unitlength}{.75pt}\put(-146,-155){
     \put(143,145)  {\scriptsize $ x^\ast $}
     \put(173,145)  {\scriptsize $ x $}
     }\setlength{\unitlength}{1pt}}
  \end{picture}}  
  = v_x : x^\ast \otimes x \rightarrow \one
  ~~,\qquad &
  \raisebox{-8pt}{
  \begin{picture}(26,22)
   \put(0,6){\scalebox{.75}{\includegraphics{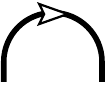}}}
   \put(0,6){
     \setlength{\unitlength}{.75pt}\put(-146,-155){
     \put(143,145)  {\scriptsize $ x $}
     \put(169,145)  {\scriptsize $ x^\ast $}
     }\setlength{\unitlength}{1pt}}
  \end{picture}}  
  = u_x^\dagger : x \otimes x^\ast \rightarrow \one
  ~~,
\\[2em]
  \raisebox{-8pt}{
  \begin{picture}(26,22)
   \put(0,0){\scalebox{.75}{\includegraphics{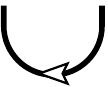}}}
   \put(0,0){
     \setlength{\unitlength}{.75pt}\put(-146,-155){
     \put(143,183)  {\scriptsize $ x $}
     \put(169,183)  {\scriptsize $ x^\ast $}
     }\setlength{\unitlength}{1pt}}
  \end{picture}}  
  = u_x : \one \rightarrow x \otimes x^\ast
  ~~,
  &
  \raisebox{-8pt}{
  \begin{picture}(26,22)
   \put(0,0){\scalebox{.75}{\includegraphics{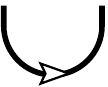}}}
   \put(0,0){
     \setlength{\unitlength}{.75pt}\put(-146,-155){
     \put(143,183)  {\scriptsize $ x^\ast $}
     \put(173,183)  {\scriptsize $ x $}
     }\setlength{\unitlength}{1pt}}
  \end{picture}}  
  = v_x^\dagger : \one \rightarrow x^\ast \otimes x~.
\end{array}
\ee
which satisfy necessary coherence properties. The quantum dimension of an
object $x$ is defined by $\dim x = v_x \circ v_x^\dagger  = u_x^\dagger \circ u_x$, 
both of which are elements of $\hom_\CC(\one,\one)=\Cb$. The quantum dimension of the category is defined by $\dim\,\CC := \sum_{i\in\mathrm{Irr}(\CC)} (\dim \, i)^2$. It is known that $\dim\,x > 0$ for $x\in \CC$ \cite[thm.\,2.3; Cor.\,2.10]{eno1}. In particular, $\dim\,\CC \geq 1$. We denote the positive square root of $\dim \CC$ by $\sqrt{\dim \CC}$.

It has a braiding structure, which amounts to an isomorphism $x\otimes y \xrightarrow{c_{x,y}} y\otimes x$ for all $x,y\in \CC$ satisfying necessary coherence conditions (i.e. hexegon relations), and we have $c_{x,y}^\dagger = c_{x,y}^{-1}$. The braiding satisfies a non-degenerate condition, which says that the so-called $S$-matrix
\be  
 s_{i,j}   ~=~  \quad \frac{1}{\sqrt{\dim \CC}}
\raisebox{-30pt}{
  \begin{picture}(110,65)
   \put(0,8){\scalebox{.75}{\includegraphics{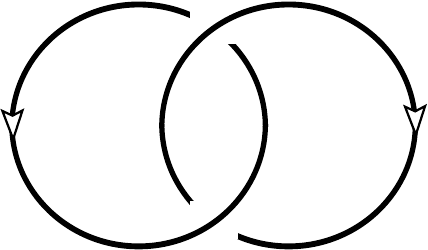}}}
   \put(0,8){
     \setlength{\unitlength}{.75pt}\put(-18,-19){
     \put( 99, 48)       {\scriptsize $ i $}
     \put( 55, 48)      {\scriptsize $ j $}
     }\setlength{\unitlength}{1pt}}
  \end{picture}}
\ee
is non-degenerate. Each simple object $i\in \mathrm{Irr}(\CC)$ has a topological spin $\theta_i \in \Cb$, which is also called a twist in mathematics. In physics, the condition $\theta_i=1$ for $i\in \mathrm{Irr}(\CC)$ amounts to say that the simple anyon $i$ is a boson.

Let us fix an orthonormal basis 
$\{ \lambda_{ij}^{k;\alpha} \}_{\alpha=1}^{N_{ij}^k}$ 
in $\hom_{\CC} (i\otimes j, k)$ and its dual basis
$\{ y^{ij}_{k;\beta} \}_{\beta=1}^{N_{ij}^k}$ 
in $\hom_{\CC} (k, i\otimes j)$, i.e. 
$$y^{ij}_{k;\beta} = (\lambda_{ij}^{k;\beta})^\dagger, \quad\quad
\lambda_{ij}^{k;\alpha} \circ y^{ij}_{k;\beta}
 = \delta_{\alpha,\beta}\, \id_{k}, \quad\quad
 \sum_{k,\beta} y^{ij}_{k;\beta} \circ \lambda_{ij}^{k;\beta} = \id_{i\otimes j}. 
$$
We denote the basis vectors graphically as follows: 
\be
\lambda_{ij}^{k; \alpha} = 
  \raisebox{-23pt}{
  \begin{picture}(30,52)
   \put(0,8){\scalebox{.75}{\includegraphics{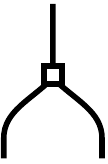}}}
   \put(0,8){
     \setlength{\unitlength}{.75pt}\put(-18,-11){
     \put(39,36)  {\scriptsize $ \alpha $}
     \put(32,61)  {\scriptsize $ k $}
     \put(17, 2)  {\scriptsize $ i $}
     \put(45, 2)  {\scriptsize $ j $}
     }\setlength{\unitlength}{1pt}}
  \end{picture}}
\quad , \qquad
y_{k; \alpha}^{ij} =
  \raisebox{-23pt}{
  \begin{picture}(30,52)
   \put(0,8){\scalebox{.75}{\includegraphics{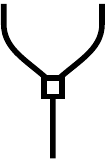}}}
   \put(0,8){
     \setlength{\unitlength}{.75pt}\put(-18,-11){
     \put(39,28)  {\scriptsize $ \alpha $}
     \put(32, 2)  {\scriptsize $ k $}
     \put(17,61)  {\scriptsize $ i $}
     \put(45,61)  {\scriptsize $ j $}
     }\setlength{\unitlength}{1pt}}
  \end{picture}} 
  ~~.
\ee
If $N_{ij}^k=1$, we abbreviate $\lambda_{ij}^{k;1}$ by $\lambda_{ij}^{k}$ and $y_{k;1}^{ij}$ by $y_k^{ij}$. We also fix 
\be \label{eq:lambda-1}
\lambda_{\one i}^{i} = l_i, \quad \lambda_{i\one}^{i} = r_i\quad \Rightarrow \quad
y_{\one i}^{i}=l_i^{-1}, \quad 
y_{i\one}^{i}=r_i^{-1} 
\ee

\begin{rem}
The simplest example of UMTC is the trivial one $\BH$, which is the category of finite dimension Hilbert spaces. It has a unique simple object (i.e. the tensor unit $\one$) given by the one dimensional Hilbert space $\Cb$. The pair $(\BH,0)$ describes the trivial 2d topological order. 
\end{rem}

\subsection{Toric Code and Ising UMTC's}

In this subsection, we give two examples of UMTC's: toric code UMTC and Ising UMTC, both of which are important to this work. 

\begin{expl} 
We denote the toric code UMTC by $\toric$. We list some of its ingredients below: 
\bnu
\item There are only four simple objects $\one, e, m, f$; 
\item The fusion rules are given by $e \otimes m = f$, $e\otimes e = m\otimes m = f\otimes f = \one$. This implies that $e^*=e$, $m^\ast=m$, $f^\ast=f$ and we have $\dim \one = \dim e = \dim m = \dim f =1$. 

\item Topological spins: $\theta_x = 1$ for $x=\one, e, m$ and $\theta_f = -1$.  

\item $S$-matrix is given by 
$$
S = \frac{1}{2} \left( \begin{array}{cccc}  1 & 1 & 1 & 1 \\ 1 & 1 & -1 & -1 \\ 1 & -1 & 1 & -1 \\ 1 & -1 & -1 & 1 \end{array} \right). 
$$
\enu
\end{expl}

\begin{rem}
The name of $\toric$ comes from the fact that the pair $(\toric, 0)$ describes the $\Zb_2$ 2d topological order, whose first lattice model realization is given by the toric code model \cite{kitaev1}. For readers with a mathematical background, $\toric$ is nothing but the Drinfeld center $Z(\rep(\Zb_2))$ of the category $\rep(\Zb_2)$ of finite dimensional representations of the group $\Zb_2$.  
\end{rem}

\begin{expl}
We denote the Ising UMTC by $\Is$. We list some of its ingredients below: 
\bnu
\item There are three simple objects $\one, \psi, \sigma$ all isomorphic to their duals, i.e. $\one=\one^\ast$, $\psi=\psi^\ast$ and $\sigma=\sigma^\ast$. Their quantum dimensions are given by $\dim \one = \dim \psi =1$ and $\dim \sigma = \sqrt{2}$.
\item The fusion rules are defined by $\psi \otimes \psi = \one$, $\psi \otimes \sigma = \sigma$, $\sigma \otimes \sigma = \one \oplus \psi$. 
\item Associators: we choose a basis such that the associators can be expressed in terms of F-matrices. In addition to (\ref{eq:lambda-1}), we further require
\be \label{eq:lambda-2}
\lambda_{\psi\psi}^\one=v_\psi, \quad y_\one^{\psi\psi} = u_\psi, \quad \lambda_{\sigma\sigma}^\one = \frac{1}{\sqrt{2}} \, v_\sigma, \quad  y_\one^{\sigma\sigma} = u_\sigma.  
\ee
There exists a choice of remaining basis, which is unique up to an arbitrary choice of $\lambda_{\psi\sigma}^{\sigma}$ (or equivalently, a choice of $\lambda_{\sigma\psi}^{\sigma}$ or $\lambda_{\sigma\sigma}^{\psi}$), realizing the following F-matrices:
\begin{align}
(\psi \otimes \sigma) \otimes \psi =\sigma &\xrightarrow{-1} \sigma = \psi \otimes (\sigma \otimes \psi) \nn
(\sigma \otimes \psi) \otimes \sigma \xrightarrow{\lambda_{\sigma\psi}^\sigma\otimes 1} \sigma \otimes \sigma \xrightarrow{\lambda_{\sigma\sigma}^\one \oplus \lambda_{\sigma\sigma}^\psi} \one \oplus \psi  &\xrightarrow{1\, \oplus \, - 1}  \one \oplus \psi \xrightarrow{y_\one^{\sigma\sigma}\oplus y_\psi^{\sigma\sigma}} \sigma \otimes \sigma \xrightarrow{1\otimes y_\sigma^{\psi\sigma}} \sigma \otimes (\psi \otimes \sigma) \nn
(\sigma \otimes \sigma) \otimes \sigma \xrightarrow{\cong} (\one \otimes \sigma) \oplus (\psi \otimes \sigma) &\xrightarrow{ \left( \begin{array}{cc} \frac{1}{\sqrt{2}} & \frac{1}{\sqrt{2}} \\ \frac{1}{\sqrt{2}} & -\frac{1}{\sqrt{2}} \end{array}\right)} (\sigma \otimes \one ) \oplus (\sigma \otimes \psi) \xrightarrow{\cong} \sigma \otimes (\sigma \otimes \sigma),
\end{align}
where we have only spelled out those non-trivial ones.

\item Braidings: 
\begin{align}
& \one \otimes x = x \xrightarrow{c_{\one, x} = 1} x = x \otimes \one \quad \text{ for } x=\one,  \psi, \sigma; \nn
& \psi \otimes \psi = \one \xrightarrow{c_{\psi, \psi} = -1} \one = \psi \otimes \psi; \nn
& \psi \otimes \sigma = \sigma \xrightarrow{c_{\psi, \sigma} = e^{-\frac{\pi i}{2}}} \sigma = \sigma \otimes \psi;  \quad\quad\quad 
\sigma \otimes \psi = \sigma \xrightarrow{c_{\sigma, \psi} = e^{-\frac{\pi i}{2}}} \sigma = \psi \otimes \sigma. \label{eq:sigma-psi} \\
& \sigma \otimes \sigma = \one \oplus \psi \xrightarrow{ c_{\sigma,\sigma} = e^{-\frac{\pi i}{8}} \, \oplus \, e^{\frac{3\pi i}{8}}} \one \oplus \psi = \sigma \otimes \sigma. \nonumber
\end{align}

\item Spins: $\theta_\one = 1, \theta_\psi = -1, \theta_\sigma = e^{\frac{\pi i}{8}}$. 

\item $S$-matrix: 
$$
S = \frac{1}{2} \left( \begin{array}{ccc}  1 & 1 & \sqrt{2} \\ 1 & 1 & -\sqrt{2} \\  \sqrt{2} & -\sqrt{2} & 0  \end{array} \right). 
$$
\enu
\end{expl}

\begin{rem}
The UMTC $\Is$ can be realized as the category of modules over the Ising vertex operator algebra $V$ with central charge $c=\frac{1}{2}$, i.e. $\Is = \Mod_V$. In particular, $V$ has three irreducible $V$-modules with the lowest conformal weights given by $0, \frac{1}{2}, \frac{1}{16}$, corresponding to three simple objects $\one, \psi, \sigma$, respectively.
\end{rem}

\begin{rem}
We use $\overline{\Is}$ to denote the same UFC as $\Is$ but with the braidings defined by the anti-braidings of $\Is$. It is called the time reverse of $\Is$. Both $\overline{\Is}$ and $\Is \boxtimes \overline{\Is}$ are UMTC's. The physical meaning of this Deligne tensor product $\boxtimes$ is the stacking of the Ising topological order with its time reversal topological order. We have $\Is \boxtimes \overline{\Is} = Z(\Is)$, where $Z(\Is)$ is the Drinfeld center of $\Is$. We will call $Z(\Is)$ the double Ising UMTC. 
\end{rem}

\section{A Gappable Gapless Edge of 2d $\Zb_2$ topological order}

In this section, we construct a gappable non-chiral gapless edge of the $\Zb_2$
topological order in three steps: in Section 3.1, we construct a gappable non-chiral
gapless edge of the double Ising topological order after a briefly review of the mathematical theory of
gapless edges established in \cite{kz2}; in Section 3.2, we construct a gapped domain wall between the double Ising topological order and the $\Zb_2$ topological order; in Section
3.3, we construct a gapless edge of the $\Zb_2$ topological order by fusing
the edge in Section 3.1 and the wall in Section 3.2.

\subsection{A Mathematical Theory of Gapless Edges} \label{sec:kz}

\begin{figure} 
$$
 \raisebox{-30pt}{
  \begin{picture}(130,130)
   \put(-20,8){\scalebox{0.6}{\includegraphics{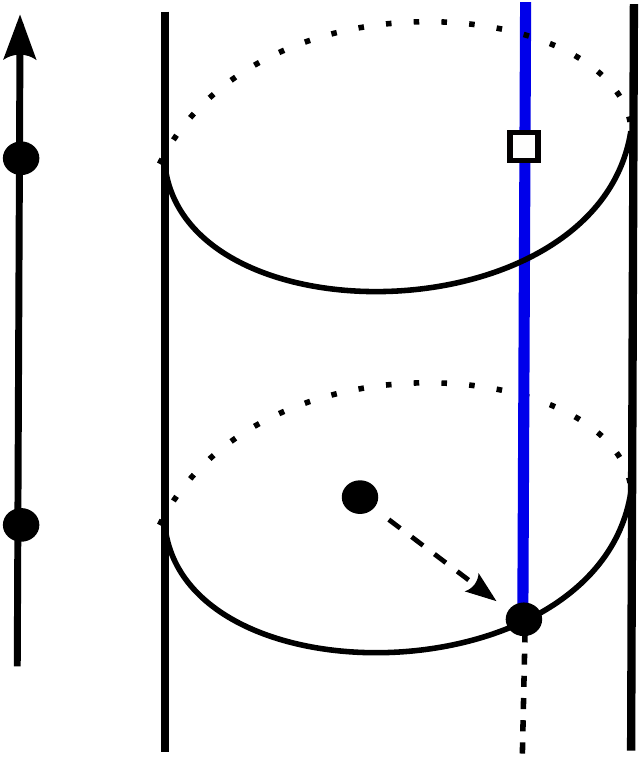}}}
   \put(-20,8){
     \setlength{\unitlength}{.75pt}\put(0,-83){
     \put(-30,133)  {$ t=0 $}
     \put(-32,219)  {$ t=t_1$}
     \put(-8, 250)  {$t$}
     \put(78,152)  {$ a \in \CC$}
     \put(126,180)  {$ A_x $}
     \put(118,262)  {$A_y$}
     \put(90,223)   {$M_{x,y}$}
     \put(75,85) {$A_{\one}=U$}
     \put(125, 105) {$x$}
     }\setlength{\unitlength}{1pt}}
  \end{picture}}
  \quad\quad\quad\quad\quad
 \raisebox{-30pt}{
  \begin{picture}(70,130)
   \put(0,5){\scalebox{0.6}{\includegraphics{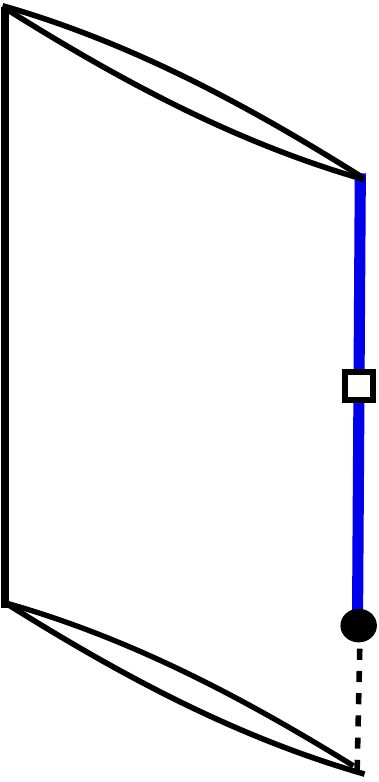}}}
   \put(0,5){
     \setlength{\unitlength}{.75pt}\put(0,-83){
     \put(88,142)  {$ A_x $}
     \put(88,205)  {$A_y$}
     \put(90,172)   {$M_{x,y}$}
     \put(88,93) {$A_{\one}=U$}
     \put(89, 115) {$x$}
     }\setlength{\unitlength}{1pt}}
  \end{picture}}
$$
$$
(a) \quad\quad\quad\quad\quad\quad\quad\quad\quad\quad\quad\quad\quad\quad\quad (b)
$$
\caption{The picture (a) depicts a 2d topological order $(\CC,c)$ on a 2-disk, together with a 1d gapless edge, propagating in time. When a topological bulk excitation $a\in \CC$ is moved to the edge at $t=0$, it creates a topological edge excitation $x$ or a boundary condition $M_x$ for the OSVOA $A_x$ living on the $t>0$ part of the world line. 
At $t=t_1>0$, the topological edge excitation $x$ is changed to another topological edge excitation $y$. This change creates a wall $M_{x,y}$ between $A_x$ and $A_y$. The picture (b) depicts the quasi-1+1D world sheet obtained by stretching the picture (a) along the dotted arrow from $a$ to $x$. 
}
\label{fig:cylinder}
\end{figure}
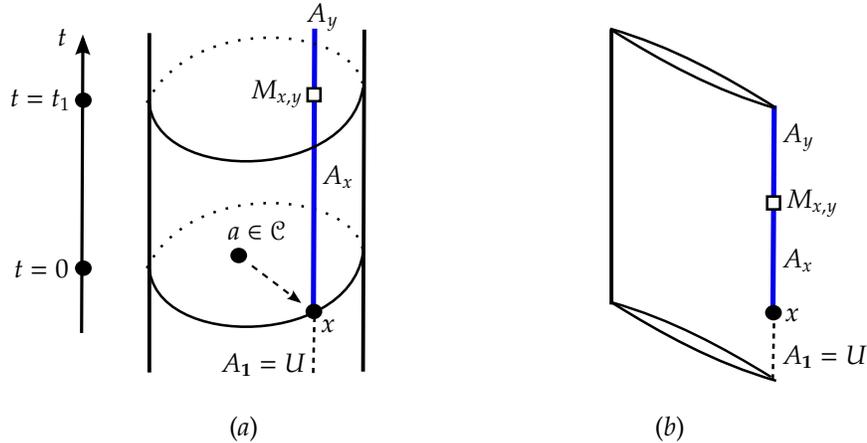

In Figure \ref{fig:cylinder} (a), we depict a spatial 2-dimensional disk propagating in time. This spatial 2-dimensional disk represents a 2d topological order $(\CC,c)$, where $\CC$ is a unitary modular tensor category of anyons and $c$ is the chiral central charge. 

If $(\CC,c)$ is a chiral topological order, there are topologically protected gapless chiral edge modes propagating on the edge of the disk, more precisely, on the 1+1D world sheet. It is well-known that these modes are states in a chiral CFT. By the state-field correspondence in a CFT, we can also say that chiral fields propagate on the 1+1D world sheet. In order to be monodromy free, these chiral fields can not contain any non-integer powers in the OPE, thus form a so-called chiral algebra $U$, or equivalently, a vertex operator algebra (VOA) in the mathematical language.

\medskip
The main idea of the mathematical theory of gapless edges established in
\cite{kz2} comes from the observation that if a bulk anyon is moved to the edge
at $t=0$, it creates a ``topological edge excitation'' (labeled by $x$) such that chiral fields living on the world line $\{ t>0 \}$ supported on $x$ (the
blue line in Figure \ref{fig:cylinder} (a)) are potentially different from those in $U$.  We denote the space of all chiral fields on this world line by $A_x$. These chiral field can have OPE along the line but no commutativity is required. Moreover, chiral fields in $A_x$ can have non-integer powers in their OPE. It turns out that chiral fields in $A_x$ have to form a boundary CFT. This can be seen from Figure\,\ref{fig:cylinder} (b), which is obtained by squeezing the solid cylinder in Figure\,\ref{fig:cylinder} (a) to a 1+1D world sheet. That $A_x$ is a boundary CFT follows from the following ``no-go theorem'':
\begin{quote}
A 1+1D boundary-bulk conformal field theory realized by a 1d lattice Hamiltonian model with boundaries should satisfy the mathematical axioms of a boundary-bulk (or open-closed) CFT of all genera \cite{huang-bcft,geometry}. In other words, it should satisfy all the modular invariant conditions and Cardy condition, etc. 
\end{quote}
For this reason, the label $x$ can also be called a {\it boundary condition}\footnote{Actually, the category of topological edge excitations is closely related to but slightly different from that of boundary conditions (for boundary CFT's) in general (see \cite{kz3}). In the case studied in this work, two categories coincide.}. We denote the trivial boundary condition by $\mathbb{1}$. It is clear  that $A_{\mathbb{1}}=U$. It is not hard to imagine that the boundary condition $x$ can be change to another one $y$ at some other point, say $t=t_1>0$, on the world line as depicted in Figure\,\ref{fig:cylinder} (a). The chiral fields living on the 0D domain wall between $A_x$ and $A_y$ are the so-called {\it boundary condition changing operators}. We denote the space of all such boundary condition changing operators by $M_{x,y}$. It is clear that $M_{x,x}=A_x$. Boundary condition changing operators can also have OPE, which defines a linear map
\be \label{eq:OPE}
M_{y,z} \otimes_\Cb M_{x,y} \to M_{x,z}. 
\ee

It was shown in \cite{kz2} that $U, A_x, M_{x,y}$ should satisfy some compatibility conditions called {\it $V$-invariant boundary condition}, where $V$ is a sub-VOA of $U$ and will be called the {\it chiral symmetry} of the edge. We assume that $V$ is unitary and rational, by which we mean that the category $\Mod_V$ of $V$-modules is a UMTC \cite{huang-mtc}. This $V$-invariant boundary condition implies that $M_{x,y}$ is a $V$-module, and the linear map in (\ref{eq:OPE}) is an intertwining operator of $V$, or equivalently, it defines a morphism $M_{y,z} \otimes_V M_{x,y} \to M_{x,z}$ in the category $\Mod_V$ of $V$-modules, where $\otimes_V$ is the tensor product in $\Mod_V$ \cite{hj}. Moreover, there is a canonical injective $V$-module map $\iota_x: V \hookrightarrow A_x$ for each $x$. Therefore, we obtain a categorical structure $\CX^\sharp$: 
\begin{itemize}
\item objects of $\CX^\sharp$ are topological edge excitations: $x,y,z, \cdots$; 
\item for each pair $(x,y)$ of objects, the hom space $\hom_{\CX^\sharp}(x, y):=M_{x,y}$ is an object in $\Mod_V$; 
\item there is an identity morphism $\iota_x: V=\one_{\Mod_V} \to A_x$ in $\Mod_V$;
\item there is a composition morphism $M_{y,z} \otimes_V M_{x,y} \to M_{x,z}$ in $\Mod_V$;
\end{itemize}
satisfying some natural conditions such that $\CX^\sharp$ is a category enriched in $\Mod_V$ or an $\Mod_V$-enriched category.

\begin{figure} 
$$
 \raisebox{-70pt}{
  \begin{picture}(80,165)
   \put(-30,8){\scalebox{0.6}{\includegraphics{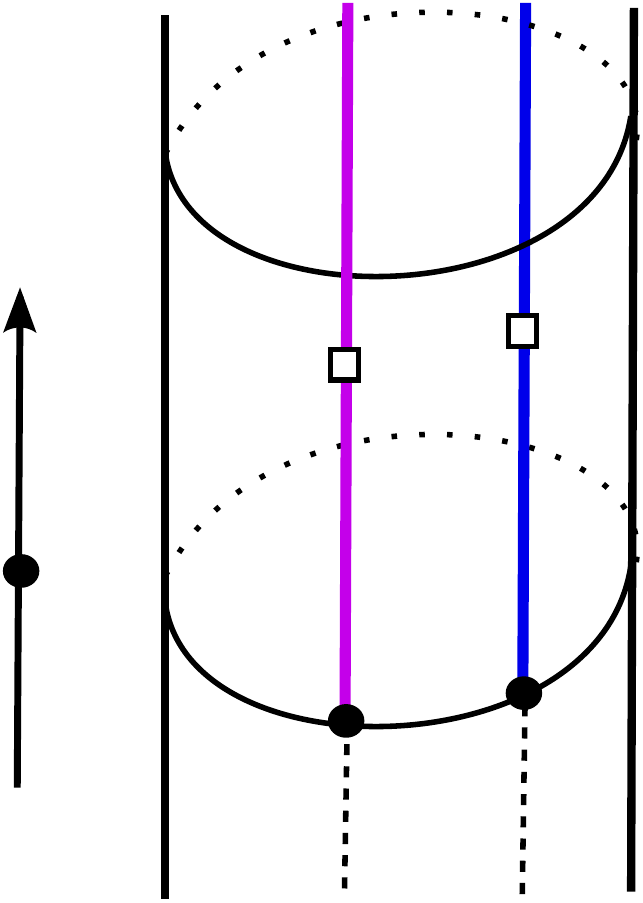}}}
   \put(-30,8){
     \setlength{\unitlength}{.75pt}\put(0,-70){
     \put(-30,140)  {$ t=0 $}
     \put(-10,180)  {$t$}
     \put(101,145)  {$ A_x $}
     \put(89,200)  {$ M_{x,y} $}
     \put(113,285)  {$A_y$}
     \put(43,192)  {$M_{x',y'}$}
     \put(59,139)  {$A_{x'}$}
     \put(73,285)  {$A_{y'}$}
     \put(65,75) {$U$}
     \put(125, 105) {$x$}
     \put(82,98)  {$x'$}
     }\setlength{\unitlength}{1pt}}
  \end{picture}}
\quad\quad\quad\quad \Rightarrow \quad\quad\quad 
 \raisebox{-70pt}{
  \begin{picture}(100,165)
   \put(10,8){\scalebox{0.6}{\includegraphics{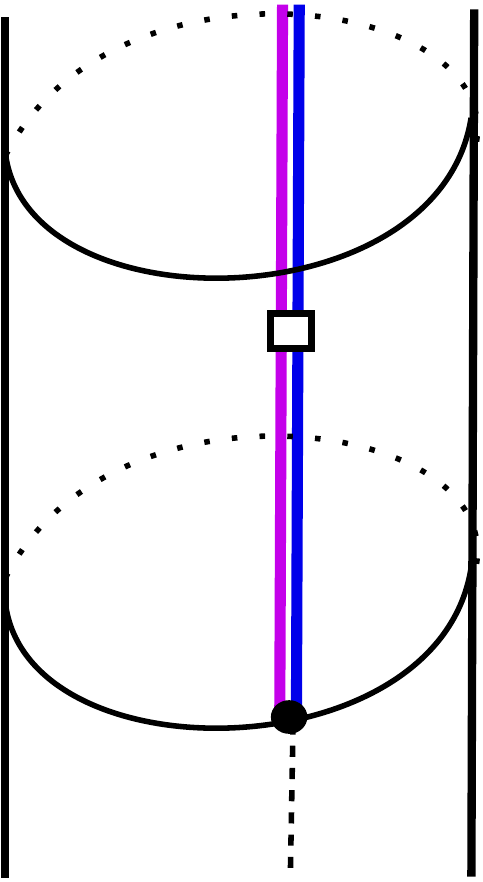}}}
   \put(10,8){
     \setlength{\unitlength}{.75pt}\put(-40,-75){
     \put(58,198)  {$M_{x'x,y'y}$}
     \put(76,145)  {$A_{x'x}$}
     \put(98,285)  {$A_{y'y}$}
     \put(93,80) {$U$}
     \put(112,102)  {$x'x$}
     }\setlength{\unitlength}{1pt}}
  \end{picture}}  
$$
$$
(a) \quad\quad\quad\quad\quad\quad\quad\quad\quad\quad\quad\quad\quad\quad\quad\quad
(b)
$$
\caption{This picture depicts a horizontal fusion of two boundary conditions, together with a horizontal fusion of $M_{x,y}$ and $M_{x',y'}$. For convenience, we abbreviate $x'\otimes x$ to $x'x$ in the picture. 
}
\label{fig:fusion}
\end{figure}
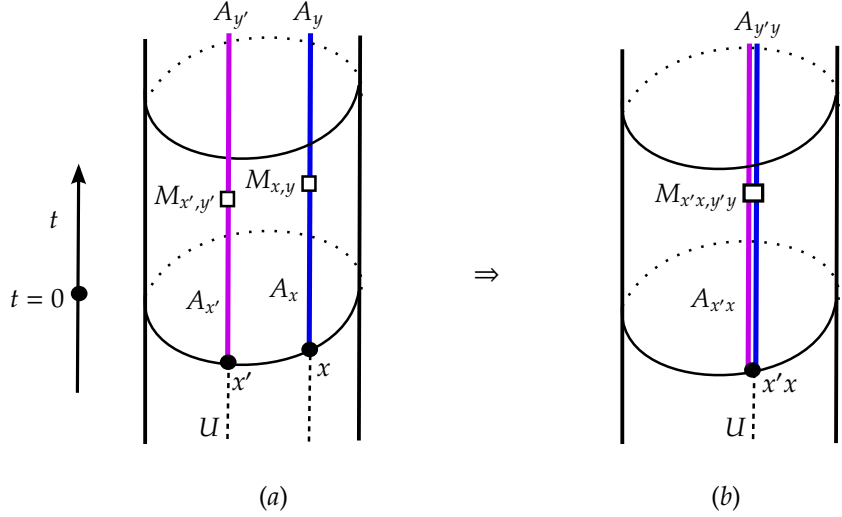

\smallskip
The last piece of structure is the horizontal fusion of topological edge excitations as depicted in Figure\,\ref{fig:fusion}, denoted by $\otimes$. It automatically provides a horizontal fusion between chiral fields on two parallel world lines. This fusion provides   
\begin{itemize}
\item a morphism $M_{x',y'}\otimes_V M_{x,y} \to M_{x'\otimes x, y'\otimes y}$ in $\Mod_V$ for objects $x,y,x',y'$ in $\CX^\sharp$, 
\end{itemize}
satisfying some natural properties. It upgrades $\CX^\sharp$ to an $\Mod_V$-enriched monoidal category \cite{mp}. We denote this gapless edge by a pair $(V, \CX^\sharp)$.

\medskip
\begin{expl} \label{expl:ising}
The well known Ising topological order $(\Is, \frac{1}{2})$ has a canonical chiral gapless edge described by a pair $(V, \Is^\sharp)$, where 
\begin{itemize}
\item $V$ is the Ising VOA of the central charge $c=1/2$ and $\Mod_V=\Is$ 
\item $\Is^\sharp$ is the $\Is$-enriched fusion category defined as follows: 
\bnu

\item objects in $\Is^\sharp$ are the same as objects in $\Is$, i.e. $x=\one, \psi, \sigma$ and their direct sums; 

\item for $x, y \in \Is$, we have the hom space defined by $\hom_{\Is^\sharp}(x,y):= M_{x,y}=y\otimes x^\ast$;  

\item the identity morphism is defined by $\one \xrightarrow{u_x} x\otimes x^\ast = M_{x,x}$; 

\item the composition morphism $M_{y,z}\otimes M_{x,y} \xrightarrow{\circ} M_{x,z}$ is defined by $z\otimes y^\ast \otimes y\otimes x^\ast \xrightarrow{1\otimes v_y \otimes 1} z\otimes x^\ast$; 

\item a fusion product between two objects is the same as the one in $\Is$; the fusion product on hom space is a morphism 
$M_{x',y'}\otimes M_{x,y} \to M_{x'\otimes x, y'\otimes y}$ defined by 
$$
y' \otimes x'^\ast \otimes y\otimes x^\ast \xrightarrow{1\otimes c_{x'^\ast, y\otimes x^\ast}} (y'\otimes y) \otimes (x' \otimes x)^\ast. 
$$
\enu
\end{itemize}

\end{expl}

The enriched monoidal category $\Is^\sharp$ is a special case of the standard construction given by Morrison and Penneys in \cite{mp}. We recall this construction only in a special setting. Let $\CB$ be a braided fusion category and $\overline{\CB}$ its time reversal. Let $\CM$ be a fusion category and $Z(\CM)$ its Drinfeld center. Let $F: \overline{\CB} \to Z(\CM)$ be a braided monoidal functor. Then we have a functor $\odot: \overline{\CB} \times \CM \to \CM$ defined by the composition of the following functors: 
$$
\overline{\CB} \times \CM \xrightarrow{F \times \id_\CM} Z(\CM) \times \CM \xrightarrow{\mathrm{forget}\times \id_\CM} \CM \times \CM \xrightarrow{\otimes} \CM,
$$
where $\mathrm{forget}: Z(\CM) \to \CM$ is the forgetful functor (by forgetting the half-braidings). 
There is a canonical construction of a $\CB$-enriched monoidal category $\CM^\sharp$ from the pair $(\CB,\CM)$, where objects in $\CM^\sharp$ are objects in $\CM$, and for $x,y\in \CM$, $\hom_{\CM^\sharp}(x,y)$ is defined by the so-called internal hom $[x,y]$ in $\overline{\CB}$ (or in $\CB$). More precisely, $[x,y]$ is uniquely determined by the following adjunction relation:
\be \label{eq:[xy]}
\hom_\CM(b \odot x, y) \simeq \hom_\CB(b, [x,y]), \quad\quad \forall b\in \CB. 
\ee
For convenience, we simply denote $\CM^\sharp$ by the pair $(\CB, \CM)$. Such a $\CB$-enriched monoidal category will be called a $\CB$-enriched fusion category.

Note that $\Is^\sharp=(\Is, \Is)$ is an $\Is$-enriched fusion category, and $M_{x,y}=y\otimes x^\ast$ is the internal hom $[x,y]$ because $[x,y]=y\otimes x^\ast$ is clearly a solution to the 
the adjunction ``equation'' (\ref{eq:[xy]}) when $\CM=\CB=\Is$. Therefore, we denote $(V, \Is^\sharp)$ by the triple $(V, \Is, \Is)$. 


\medskip
We give an example of gappable non-chiral gapless edges. 
\begin{expl} \label{expl:double-ising}
By folding a Ising topological order $(\Is,\frac{1}{2})$ with the canonical chiral gapless edge, we obtain a double Ising topological order $(\Is \boxtimes \overline{\Is}, 0)$ equipped with both chiral gapless edge modes described by $(V, \Is, \Is)$ and anti-chiral gapless edge modes described by $(\overline{V}, \overline{\Is}, \overline{\Is})$, where $\overline{V}$ is the same VOA as $V$ but contains only anti-chiral fields $\phi(\bar{z}), \forall \phi \in V$. Altogether, they form a non-chiral gapless edge of the double Ising topological order $(\Is \boxtimes \overline{\Is}, 0)$ described by the triple:
\be \label{eq:double-ising}
(V\otimes_\Cb \overline{V}, \,\, \Is \boxtimes \overline{\Is}, \,\, \Is \boxtimes \overline{\Is})).
\ee
This non-chiral gapless edge is clearly gappable. 
\end{expl}

\subsection{A Gapped Wall between Double Ising and Toric Code} \label{sec:wall}

It was first shown by Bais and Slingerland in \cite{bs} on a physical level of rigor that one can obtain the $\Zb_2$ topological order from the double Ising via an anyon condensation. In this subsection, we will give a complete and rigorous derivation of this result based on the mathematical theory of anyon condensation developed in \cite{kong-anyon}. As a byproduct, we construct a gapped domain wall between the double Ising and the $\Zb_2$ topological orders.

\medskip
An anyon condensation from an old topological order $(\CC,c)$ to a new phase $(\CD,c)$ is controlled by a condensable algebra $A$ in $\CC$. We recall this notion below. 
\begin{defn} \label{def:alg} {\rm
Let $\CC$ be a UMTC. An algebra $A$ in $\CC$ is a triple $(A, \mu, \iota)$, where $A$ is an object in $\CC$, $\mu:A\otimes A\to A$ and $\iota: \one \to A$ are morphisms satisfying the following conditions: 
$$
\mu \circ (\mu \otimes \id_A) \circ \alpha_{A,A,A} = \mu \circ (\id_A \otimes \mu).
$$
$$
\mu \circ (\iota \otimes \id_A) = \id_A = \mu\circ (\id_A \otimes \iota). 
$$
The algebra $A$ is called commutative if $\mu = \mu \circ c_{A,A}$. }
\end{defn}

\begin{defn} \label{def:module} {\rm
A right $A$-module in $\CC$ is a pair $(M, \mu_M)$, where $M$ is an object in $\CC$ and $\mu_M: M\otimes A \to M$ is such that 
$$
\mu_M \circ (\id_M \otimes \mu) = \mu_M \circ (\mu_M \otimes \id_A) \circ \alpha_{M,A,A}, \quad\quad \mu_M \circ (\id_M\otimes \iota) =\id_M. 
$$
A left module and a bimodule can be defined similarly.}
\end{defn}

\begin{defn} \label{def:separable} {\rm
An algebra $(A, \mu, \iota)$ is called {\it separable} if $\mu: A\otimes A \to A$ splits as a morphism of $A$-$A$-bimodule. Namely, there is an $A$-$A$-bimodule map $e: A\to A\otimes A$ such that $\mu \circ e = \id_A$. 
A separable algebra is called {\it connected} if $\dim \hom_\CC(\one, A) = 1$. We will call a connected commutative separable algebra as a {\it condensable algebra}. A condensable algebra $A$ is called {\it Lagrangian} if $\dim(A)^2=\dim(\CC)$.

}
\end{defn}


A right $A$-module is called a {\it local $A$-module} if $\mu_M\circ c_{A,M} \circ c_{M,A}=\mu_M$. If $A$ is condensable, the category $\CC_A$ of right $A$-modules is a UFC. Its fusion sub-category $\CC_A^0$ consisting of local $A$-modules in $\CC$ is a UMTC with the braidings, rigidity and spins inherited from those in $\CC$. In particular, for a local $A$-module
$\hat{M}=(M, \mu_{\hat{M}})$, the spin (or twist) $\theta_{\hat{M}}^A$ in
$\CC_A^0$ is defined by $\theta_M$ in $\CC$. Moreover, we have the following
identities \cite{KO}: \be \label{eq:ZC_A} \dim(\CC_A) =
\frac{\dim(\CC)}{\dim(A)}, \quad\quad \dim(\CC_A^0) = \frac{\dim(\CC)}{\dim(A)^2},
\quad\quad \dim_A (x) = \frac{\dim x}{\dim A}, \ee where $\dim\, x$ is the
quantum dimension of $x$ in $\CC$ and $\dim_A(x)$ is that of $x$ in $\CC_A$. If, in addition, $A$ is Lagrangian, we have
$\CC_A^0=\BH$. 

\begin{expl}
There are two condensable algebras in $\toric$: $A_e=\one \oplus e$ and $A_m=\one \oplus m$. Both of them are Lagrangian. 
\end{expl}

\begin{expl}
Since a condensable algebra $A$ satisfies $\theta_A =\id_A$, the only condensable algebra $A$ in $\Is$ is $\one$. 
\end{expl}

\begin{expl}
Let $Z(\Is) = \Is \boxtimes \overline{\Is} \xrightarrow{\otimes} \Is$ be the tensor product functor and let $R: \Is \to Z(\Is)$ be its right adjoint functor. 
\bnu
\item $R(\one)=\oplus_{i\in \mathrm{Irr}(\Is)} i^\ast \boxtimes i$ has the structure of a Lagrangian algebra naturally induced from the algebraic structure on $\one$ \cite{kr}. More explicitly, the multiplication map is defined by (first appeared in \cite{mueger,ffrs}, we take it from \cite[Eq.\,(2.58)]{kr}): 
$$
\bigoplus_{i,j,k\in \mathrm{Irr}(\CC)} \sum_{\alpha} ~~
  \raisebox{-25pt}{
  \begin{picture}(70,50)
   \put(0,8){\scalebox{.75}{\includegraphics{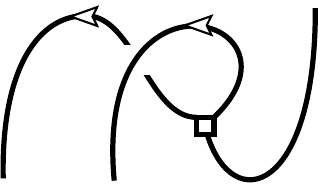}}}
   \put(0,8){
     \setlength{\unitlength}{.75pt}\put(-18,-11){
     \put(19, 2)        {\scriptsize $ i^\ast $}
     \put(48, 2)        {\scriptsize $ j^{\ast} $}
     \put(108, 65)  {\scriptsize $k^{\ast}$}
     \put(74, 35)      {\scriptsize $\alpha$}
     }\setlength{\unitlength}{1pt}}
  \end{picture}}
~~  \boxtimes ~~
\raisebox{-25pt}{
  \begin{picture}(20,50)
   \put(0,8){\scalebox{.75}{\includegraphics{pic-lambda-eps-converted-to.pdf}}}
   \put(0,8){
     \setlength{\unitlength}{.75pt}\put(-18,-11){
     \put(18, 2)     {\scriptsize $ i $}
     \put(46, 2)     {\scriptsize $ j $}
     \put(30, 26)   {\scriptsize $\alpha$} 
     \put(32, 62) {\scriptsize $ k $ }
     }\setlength{\unitlength}{1pt}}
  \end{picture}}
  \phantom{\frac{\dim U_i \,\dim U_j}{\dim U_k \,\dim\,\CC}} 
$$
and the unit map is defined by the canonical embedding $\one_{Z(\Is)} \hookrightarrow R(\one)$. 

\item We denote the fusion sub-category of $\Is$ consisting of $\one$ and $\psi$ by $\CA$. Then $\CA \boxtimes \overline{\CA}$ is a fusion sub-category of $Z(\Is)$. We have a subalgebra $A$ of $R(\one)$ defined by:
$$
A:= R(\one) \cap (\CA \boxtimes \overline{\CA}) =\one\boxtimes \one \oplus \psi \boxtimes \psi
$$
which is also a condensable algebra in $\CA \boxtimes \overline{\CA}$. More explicitly, we can write down the multiplication map of $A$ as follows: 
\begin{align}
&\left( \one \otimes \one \xrightarrow{r_\one} \one \right) \boxtimes  \left( \one \otimes \one \xrightarrow{r_\one} \one \right) \quad \oplus \quad
\left( \one \otimes \psi \xrightarrow{r_\psi} \psi \right) \boxtimes  \left( \one \otimes \psi \xrightarrow{r_\psi} \psi \right) \nn
& \left( \psi \otimes \one \xrightarrow{r_\psi} \psi \right) \boxtimes  \left( \psi \otimes \one \xrightarrow{r_\psi} \psi \right) \quad \oplus \quad 
\left( \psi \otimes \psi \xrightarrow{-v_\psi} \one \right) \boxtimes  \left( \psi \otimes \psi \xrightarrow{v_\psi} \one \right) \nonumber
\end{align}
Note that $\dim A = 2$. 
\enu
\end{expl}

\medskip
According to the anyon condensation theory \cite{kong-anyon}, by condensing the condensable algebra $A$ in the initial phase $(Z(\Is),0)$, we obtain 
\bnu
\item a new topological order $(Z(\Is)_A^0,0)$ 

\item and a 1d gapped domain wall described by the UFC $Z(\Is)_A$. 

\enu

Our goal is to workout these two categories $Z(\Is)_A$ and $Z(\Is)_A^0$ explicitly and prove that $Z(\Is)_A^0\simeq \toric$ as UMTC's. 
Before we do that, we first collect a few relevant results.

\bnu
\item Notice that we have 
\be \label{eq:q-dim}
\dim Z(\Is)_A = \frac{Z(\Is)}{\dim A}=8, \quad\quad\quad \dim Z(\Is)_A^0 = \frac{Z(\Is)}{(\dim A)^2} = \frac{16}{2^2} = 4. 
\ee

\item $A$ is clearly a simple right $A$-module and a local $A$-module. 

\item Another obvious simple right $A$-module is $(\sigma\boxtimes \sigma)$ because the following splitting
$$
R(\one) = A \oplus (\sigma\boxtimes \sigma)
$$
is a splitting of $A$-$A$-bimodules because $A$ is separable. The right $A$-module structure on $(\sigma\boxtimes \sigma)$ can be explictly defined by 
\be \label{eq:mu-A}
\left( \sigma \otimes \one \xrightarrow{r_\sigma} \sigma \right) \boxtimes  \left( \sigma \otimes \one \xrightarrow{r_\sigma} \sigma \right)
\quad \oplus \quad
\left( \sigma \otimes \psi \xrightarrow{\lambda_{\sigma\psi}^{\sigma}} \sigma \right) \boxtimes  \left( \sigma \otimes \psi \xrightarrow{\lambda_{\sigma\psi}^{\sigma}} \sigma \right). 
\ee
Using Eq.\,(\ref{eq:sigma-psi}), it is easy to see that $\sigma\boxtimes\sigma$ is a local $A$-module. 

\item The condensable algebra $A$ has a non-trivial algebraic automorphism defined by 
$$
\delta: A=(\one \boxtimes \one) \oplus (\psi \boxtimes \psi) \xrightarrow{1 \oplus -1} 
(\one \boxtimes \one) \oplus (\psi \boxtimes \psi)=A. 
$$
And $\delta$ is an involution, i.e. $\delta^2=\id_A$. 

\item We can use $\delta$ to twist the $A$-action on a right $A$-module $M=(M,\mu_M)$ and obtain a new right $A$-module structure, denoted by $M^{\mathrm{tw}}$, with a new action defined by
$$
M \otimes A \xrightarrow{ \id_M \otimes \delta} M \otimes A \xrightarrow{\mu_M} M. 
$$

\bnu
\item If $M=x\otimes A$ for $x\in\CC$ with the right $A$-action defined 
\be \label{eq:x-A-A}
x\otimes A\otimes A \xrightarrow{1\otimes \mu} x \otimes A,
\ee 
then we have $(x\otimes A) \simeq (x\otimes A)^{\mathrm{tw}}$ as right $A$-modules with the isomorphism given by $x\otimes A \xrightarrow{\id_x \otimes \delta} x\otimes A$. 
\item For $M=(\sigma\boxtimes \sigma)$, we obtain a new local $A$-module $(\sigma\boxtimes \sigma)^{\mathrm{tw}}$, which is defined by
$$
\left( \sigma \otimes \one \xrightarrow{r_\sigma} \sigma \right) \boxtimes  \left( \sigma \otimes \one \xrightarrow{r_\sigma} \sigma \right)
\quad \oplus \quad
\left( \sigma \otimes \psi \xrightarrow{-\lambda_{\sigma\psi}^{\sigma}} \sigma \right) \boxtimes  \left( \sigma \otimes \psi \xrightarrow{\lambda_{\sigma\psi}^{\sigma}} \sigma \right). 
$$
It is clear that $(\sigma\boxtimes \sigma)^{\mathrm{tw}}$ is not isomorphic to $(\sigma\boxtimes \sigma)$ as local $A$-modules. 
\enu
\enu

We need find all right $A$-modules. We use the fact that all simple objects in $Z(\Is)_A$ can be realized by a direct summand of $x\otimes A$ for $x\in \mathrm{Irr}(Z(\Is))$ with the right $A$-action defined by (\ref{eq:x-A-A}) because $x\simeq x\otimes_A A$ and $A$ is separable. Therefore, we obtain a complete list of simple right $A$-modules as follows:

\bnu

\item $(\one\boxtimes\one) \otimes A = A$. 

\item $(\psi \boxtimes\one) \otimes A = \psi \boxtimes \one \oplus \one\boxtimes \psi$ is clearly a simple local $A$-module. It is useful to write the right $A$-action on $\psi \boxtimes \one \oplus \one\boxtimes \psi$ explicitly as follows: 
\be \label{eq:A-action-1}
(r_\psi \boxtimes r_\one) \oplus (v_\psi \boxtimes l_\psi) \oplus (r_\one\boxtimes r_\psi) \oplus (- l_\psi \boxtimes v_\psi). 
\ee

\item $(\one \boxtimes\psi) \otimes A = \psi \boxtimes \one \oplus \one\boxtimes \psi$ is a simple local $A$-module. It is useful to write the right $A$-action on $\psi \boxtimes \one \oplus \one\boxtimes \psi$ explicitly as follows:
\be \label{eq:A-action-2}
(r_\psi \boxtimes r_\one) \oplus (-v_\psi \boxtimes l_\psi) \oplus (r_\one\boxtimes r_\psi) \oplus (l_\psi \boxtimes v_\psi). 
\ee
Using the explicit right $A$-actions given in Eq. (\ref{eq:A-action-1}) and (\ref{eq:A-action-2}), it is obvious to see that 
$$
(\one \boxtimes\psi) \otimes A = ((\psi \boxtimes\one) \otimes A)^{\mathrm{tw}} \simeq (\psi \boxtimes\one) \otimes A. 
$$ 
as right $A$-modules. 

\item $(\psi\boxtimes \psi) \otimes A$ is a simple local $A$-module. By writing out the right $A$-action on $\one\boxtimes\one \oplus \psi\boxtimes\psi$ in basis explicitly and comparing it with Eq. (\ref{eq:mu-A}), we see that $(\psi\boxtimes \psi) \otimes A=A^{\mathrm{tw}} \simeq A$. 

\item $(\sigma\boxtimes\sigma)\otimes A$ is a local $A$-module but not simple. By checking the right $A$-action in basis, we obtain that
$$
(\sigma\boxtimes\sigma)\otimes A \simeq (\sigma\boxtimes\sigma) \oplus (\sigma\boxtimes\sigma)^{\mathrm{tw}}
$$
as local $A$-modules. 

\item $(\sigma\boxtimes\one)\otimes A = (\sigma \boxtimes \one) \oplus (\sigma\boxtimes\psi)$ is a simple right $A$-module but not local. 

\item $(\sigma\boxtimes\psi)\otimes A = (\sigma \boxtimes \one) \oplus (\sigma\boxtimes\psi)$ is a simple right $A$-module but not local. By writing out the right $A$-action on $(\sigma \boxtimes \one) \oplus (\sigma\boxtimes\psi)$ in basis explicitly for both case 6 and 7, one see that $(\sigma\boxtimes\psi)\otimes A=((\sigma\boxtimes\one)\otimes A)^{\mathrm{tw}}\simeq (\sigma\boxtimes\one)\otimes A$. 

\item $(\one\boxtimes\sigma)\otimes A = (\one \boxtimes \sigma) \oplus (\psi\boxtimes\sigma)$ is a simple right $A$-module but not local.

\item $(\psi\boxtimes\sigma)\otimes A = (\one \boxtimes \sigma) \oplus (\psi\boxtimes\sigma)$ is a simple right $A$-module but not local.
By writing out the right $A$-action on $(\one \boxtimes \sigma) \oplus (\psi\boxtimes\sigma)$ in basis explicitly for both case 8 and 9, one can see that $(\psi\boxtimes\sigma)\otimes A=((\one\boxtimes\sigma)\otimes A)^{\mathrm{tw}}\simeq (\sigma\boxtimes\one)\otimes A)$. 

\enu
To summarize, we have found all simple objects in $Z(\Is)_A$: 
\begin{itemize}
\item four simple local $A$-modules with new and shorter notations: 
\be \label{eq:1emf}
\mathbb{1}:=A=(\one\boxtimes \one) \oplus (\psi \boxtimes \psi), \quad e:=(\sigma\boxtimes\sigma), \quad m:=(\sigma\boxtimes\sigma)^{\mathrm{tw}}, \quad
f:= (\psi\boxtimes \one) \otimes A=\psi \boxtimes \one \oplus \one\boxtimes \psi
\ee
with quantum dimensions in $Z(\Is)_A$ all given by $1$. 
\item two simple non-local right $A$-modules with new and shorter notations:
\be \label{eq:chi}
 \chi_+:=(\one\boxtimes\sigma)\otimes A = (\one \boxtimes \sigma) \oplus (\psi\boxtimes\sigma) \quad \quad 
 \chi_-:=(\sigma\boxtimes\one)\otimes A = (\sigma\boxtimes\one) \oplus (\sigma\boxtimes\psi)
\ee
with quantum dimensions in $Z(\Is)_A$ both given by $\sqrt{2}$. 
\end{itemize}

\begin{rem}
One can also check directly from the sum of quantum dimensions and Eq. (\ref{eq:q-dim}) to see that we have found all simple right $A$-modules and all simple local $A$-modules. 
\end{rem}

Now we work out the fusion rules of $Z(\Is)_A^0$. Note that the fusion product in $Z(\Is)_A^0$ is given by the relative tensor product $\otimes_A$. In order to see how $Z(\Is)_A^0$ can be identified with the UMTC $\toric$ as an abstract tensor category. We adopt the new convention of notation as illustrated by the following examples: 
$$
e \star m := (\sigma \boxtimes \sigma) \otimes_A (\sigma \boxtimes \sigma)^{\mathrm{tw}}, \quad
\chi_+ \star f := [(\one \boxtimes\sigma)\otimes A]\otimes_A [(\psi\boxtimes \one) \otimes A].
$$ 
where we have also replaced the tensor product $\otimes_A$ in $Z(\Is)_A$ by $\star$. 
\bnu
\item The right $A$-module structure on $(\sigma\boxtimes\sigma)$ gives a left $A$-module structure on $(\sigma\boxtimes\sigma)^\ast=(\sigma\boxtimes\sigma)$. One can check that it coincides with the left $A$-module structure on $(\sigma\boxtimes\sigma)$ defined by 
$$
A \otimes (\sigma\boxtimes\sigma) \xrightarrow{c_{A, \sigma\boxtimes\sigma}} (\sigma\boxtimes\sigma) \otimes A \to A. 
$$
As a consequence, the dual of $(\sigma\boxtimes\sigma)$ in $Z(\Is)_A^0$ is precisely $(\sigma\boxtimes\sigma)$. In other words, we must have $(\sigma\boxtimes\sigma)\otimes_A (\sigma\boxtimes\sigma) \simeq A$ as right $A$-modules. 

\item Similarly, we have 
$(\sigma\boxtimes\sigma)^{\mathrm{tw}} \otimes_A (\sigma\boxtimes\sigma)^{\mathrm{tw}} \simeq A$ as right $A$-modules. 

\item On the one hand, we have 
$$
[(\sigma\boxtimes \sigma)\otimes A] \otimes_A (\sigma\boxtimes \sigma) \simeq [(\sigma\boxtimes \sigma) \oplus (\sigma\boxtimes \sigma)^{\mathrm{tw}} ] \otimes_A (\sigma\boxtimes \sigma). 
$$
On the other hand, we have 
$$
[(\sigma\boxtimes \sigma)\otimes A] \otimes_A (\sigma\boxtimes \sigma) \simeq (\sigma\boxtimes \sigma) \otimes (\sigma\boxtimes \sigma)
\simeq (\one\oplus\psi)\boxtimes (\one\oplus\psi) \simeq A \oplus  [(\psi \boxtimes \one)\oplus (\one\boxtimes \psi)]. 
$$
Therefore, we obtain 
$(\sigma\boxtimes \sigma)^{\mathrm{tw}} \otimes_A (\sigma\boxtimes \sigma) \simeq  (\psi\boxtimes\one)\otimes A$ as right $A$-modules. 
\enu
As a consequence, we obtain the following fusion rules of $Z(\Is)_A^0$: 
\be \label{eq:fusion-rule-1}
\mathbb{1}\star x = x, \quad e\star e = m\star m = f\star f = \mathbb{1}, \quad e\star m = f, 
\ee
where $x=\mathbb{1},e,m,f$. 

\medskip
Note that this fusion rule coincides with that of $\toric$. According to \cite{rsw}, there are exactly two modular tensor categories shares these fusion rules. They are $\toric$ and $\mathrm{Spin}(8)_1$, the spins of which are given by $(1,1,1,-1)$ and $(1,-1,-1,-1)$, respectively. In our case, we have
$$
\theta_e^A = \theta_m^A = \theta_\sigma \cdot \theta_\sigma^{-1} = 1, \quad\quad \theta_f^A= \theta_\psi=-1,
$$
where $\theta_-^A$ denotes the spins in $Z(\Is)_A^0$. 
Therefore, we have proved the following result. 

\begin{thm}
$Z(\Is)_A^0 \simeq \toric$ as UMTC's. 
\end{thm}

It is useful to work out the remaining fusion rules in $Z(\Is)_A$ as follows: 
\be \label{eq:fusion-rule-2}
\chi_{\pm} \star \chi_{\pm} = \mathbb{1}\oplus f, \quad \quad \chi_{\pm} \star \chi_{\mp} = e\oplus m, \quad \quad 
e\star \chi_{\pm} = \chi_{\pm} \star e = m\star \chi_{\pm} = \chi_{\pm} \star m = \chi_{\mp}, 
\ee
We give a proof below. 
\bnu

\item $\chi_+ \star \chi_+= [(\one \boxtimes \sigma) \otimes A] \otimes_A [(\one\boxtimes \sigma) \otimes A] \simeq (\one\boxtimes\sigma)\otimes(\one\boxtimes\sigma) \otimes A \simeq A \oplus (\one\boxtimes\psi) \otimes A = \mathbb{1} \oplus f$. 

\item $\chi_- \star \chi_- = [(\sigma \boxtimes \one) \otimes A] \otimes_A [(\sigma\boxtimes \one) \otimes A] \simeq (\sigma\boxtimes\one)\otimes(\sigma\boxtimes\one) \otimes A \simeq A \oplus (\psi \boxtimes\one) \otimes A = \mathbb{1}\oplus f$. 

\item $\chi_+ \star \chi_-=[(\one \boxtimes \sigma) \otimes A] \otimes_A
  [(\sigma \boxtimes \one) \otimes A] \simeq (\sigma\boxtimes\sigma)\otimes A = e\oplus m = \chi_- \star \chi_+$. 

\item $e\star \chi_+ \simeq \chi_+\star e =[(\one\boxtimes\sigma)\otimes A] \otimes_A  (\sigma \boxtimes \sigma) \simeq \sigma \boxtimes (\one\oplus \psi) = \chi_-$. 

\item $e\star \chi_- \simeq \chi_-\star e=[(\sigma\boxtimes\one)\otimes A] \otimes_A  (\sigma \boxtimes \sigma) \simeq (\one\oplus \psi) \boxtimes \sigma = \chi_+$. 

\item $m\star \chi_+ \simeq \chi_+ \star m =[(\one\boxtimes\sigma)\otimes A] \otimes_A (\sigma \boxtimes \sigma)^{\mathrm{tw}} \simeq \chi_-$. 

\item $m\star \chi_- \simeq \chi_- \star m =[(\sigma\boxtimes\one)\otimes A] \otimes_A (\sigma \boxtimes \sigma)^{\mathrm{tw}} \simeq \chi_+$. 
\enu

\begin{rem}
From above identities, one can see that $x=x^\ast$ and $x\star y=y\star x$ as objects for all $x\in Z(\Is)_A$.
\end{rem}

In summary, by condensing $A$ in the double Ising topological order $(Z(\Is),0)$, we obtain the 2d $\Zb_2$ topological order and a gapped domain wall given by the UFC $Z(\Is)_A$, which consists of six simple wall excitations $\mathbb{1}, e, m, f, \chi_{\pm}$.

\subsection{A Gappable Non-chiral Gapless Edge of $\Zb_2$ topological order} \label{sec:edge}

In this subsection, we combine results in Section\,\ref{sec:kz} and \ref{sec:wall} to construct a gappable non-chiral gapless edge of the $\Zb_2$ topological order.

\medskip
In Example\,\ref{expl:double-ising}, we have constructed a gappable non-chiral gapless edge of the double Ising topological order. It can be expressed by a triple 
\be \label{eq:gapless-edge}
(V\otimes_\Cb \overline{V}, \,\, Z(\Is), \,\, Z(\Is)).
\ee

In Section\,\ref{sec:wall}, we have constructed a gapped domain wall $Z(\Is)_A$ between the double Ising and the $\Zb_2$ topological orders. Using the notion of enriched fusion category introduced in Section\,\ref{sec:kz}, the UFC $Z(\Is)_A$ can also be viewed as an enriched fusion category determined by the pair $(\BH, Z(\Is)_A)$. Therefore, this gapped domain wall can also be expressed as a triple 
\be \label{eq:gapped-wall}
(\Cb, \,\, \BH, \,\, Z(\Is)_A), 
\ee
where the complex numbers $\Cb$ should be viewed as the trivial VOA of central charge $0$.

\begin{figure}[htbp] 
$$
 \raisebox{-40pt}{
  \begin{picture}(120,80)
   \put(-50,8){\scalebox{0.8}{\includegraphics{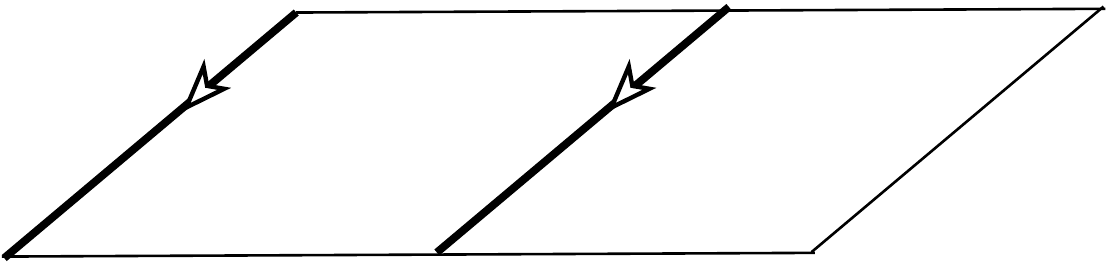}}}
   \put(-50,8){
     \setlength{\unitlength}{.75pt}\put(0,0){
     \put(-30,85)  {$ (V\otimes_\Cb \overline{V}, Z(\Is), Z(\Is)) $}
     \put(95,40)  {$ Z(\Is) $}
     \put(220,25)  {$\toric$}
     \put(210,85)  {$(\Cb, \BH, Z(\Is)_A)$}
     \put(-25,-10)  {a gapless edge}
     \put(110,-10) {a gapped wall}
     }\setlength{\unitlength}{1pt}}
  \end{picture}}
$$
\caption{This picture depicts a gapped domain wall between the double Ising and the $\Zb_2$ topological order, and a gapless edge of the double Ising topological order.    
}
\label{fig:edge-wall}
\end{figure}
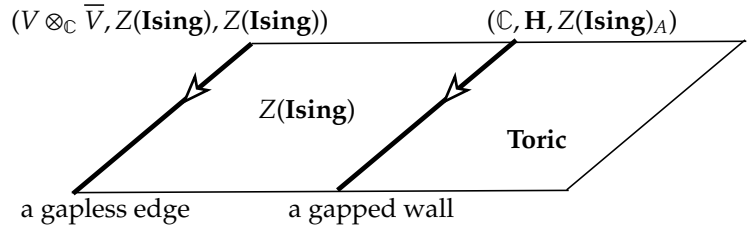

By fusing the gapless edge (\ref{eq:gapless-edge}) and the gapped domain wall (\ref{eq:gapped-wall}) as depicted in Figure\,\ref{fig:edge-wall}, we obtain a non-chiral gapless edge of the $\Zb_2$ topological order. 
Using the fusion product defined in \cite[Eq.\,(5.2)]{kz2}, this gapless edge is given by the following triples: 
\begin{align}
&(V\otimes_\Cb \overline{V}, \,\, Z(\Is), \,\, Z(\Is)) \boxtimes_{(Z(\Is),0)} (\Cb, \,\, \BH, \,\, Z(\Is)_A) \nn
&\hspace{2cm} = (V\otimes_\Cb \overline{V} \otimes_\Cb \Cb, \,\, Z(\Is) \boxtimes \BH, \,\, Z(\Is)\boxtimes_{Z(\Is)} Z(\Is)_A) \nn
&\hspace{2cm} = (V\otimes_\Cb \overline{V}, \,\, Z(\Is), \,\, Z(\Is)_A).
\end{align}
The validity of this fusion formula will be explained in details in \cite{kz3}. We can describe the observables on this gapless edge $(V\otimes_\Cb \overline{V}, \,\, Z(\Is), \,\, Z(\Is)_A)$ explicitly as follows :
\begin{itemize}

\item topological edge excitations (or boundary conditions) on this gapless edge are objects in $Z(\Is)_A$. In particular, there are exactly six simple ones: $\mathbb{1}, e, m, f, \chi_{\pm}$. They can be fused horizontally according to the fusion rules in $Z(\Is)_A$ (see Eq.\,(\ref{eq:fusion-rule-1}) and (\ref{eq:fusion-rule-2})). 

\item Boundary CFT's and walls between them $M_{x,y}$ are given by internal homs $[x,y]=(x \otimes_A y^\ast)^\ast$ \cite{ostrik}.  We want to work out each $[x,y]$ as objects in order to compare them with the partition functions obtained in lattice model realizations in Section\,\ref{sec:lattice}. Recall that $x^\ast=x$ for $x\in Z(\Is)_A$. Moreover, for $x,y=\mathbb{1},e,m,f,\chi_{\pm}$, 
we obtain
\be \label{eq:partition-function-1}
M_{x,y}= [x,y]= [\mathbb{1}, x\otimes_A y] = x\otimes_A y, \quad \mbox{as objects}.
\ee
Each $x$ is defined by objects in $Z(\Is)$ by Eq.\,(\ref{eq:1emf}) and (\ref{eq:chi}). 
\end{itemize}
This non-chiral gapless edge is clearly gappable.

\begin{rem}
In this work, we only scratch the surface of gappable non-chiral gapless edges. A detailed theory of gapless edges will be provided in \cite{kz3,kz4}. 
\end{rem}

\subsection{The partition functions of $M_{x,y}$}

Recall that the 2d $\Zb_2$ topological order has two gapped edges. One is obtained by condensing $m$-particles and the other by condensing $e$-particles \cite{bk}. Both gapped edges can be described by the same UFC $\rep(\Zb_2)$ but equipped with two different bulk-to-edge functors $\toric \to \rep(\Zb_2)$, corresponding to different condensations. The bulk-edge correspondence says that the Drinfeld center of $\rep(\Zb_2)$ gives the bulk, i.e. bulk = the center of the edge. A pure edge topological phase transition between these two gapped edges closes the gap and produces a gappable gapless edge at the critical point. Since the bulk-edge correspondence holds before and after the transition, we expect that it holds at the critical point as well.

In Section\,\ref{sec:edge}, we have constructed a gappable gapless edge $(V\otimes_\Cb \overline{V}, \,\, Z(\Is), \,\, Z(\Is)_A)$ of the 2d $\Zb_2$ topological order.  It shares the same bulk with the two gapped edges. Mathematically, it is because that the enriched fusion category $(Z(\Is), \, Z(\Is)_A)$ and the UFC $\rep(\Zb_2)$ share the same Drinfeld center \cite{kz1}, or equivalently, they are Morita equivalent as enriched fusion categories \cite{zheng}. Physically, it is because one can construct a gapless 0d domain walls between this gappable gapless edge and each of the two gapped edges. The physical reason is equivalent to the mathematical one. We will give more details  about this in \cite{kz3,kz4}.

Therefore, it is reasonable to ask if this gappable gapless edge describes the critical point of a pure edge topological phase transition between two gapped edges. The main goal of this work is to prove that this is indeed true. We will achieve this goal in Section\,\ref{sec:lattice} by recovering the chiral symmetry $V$ and all the ingredients of the enriched fusion category $(Z(\Is), \,\, Z(\Is)_A)$, i.e. $M_{x,y}$, from a lattice model construction.

\begin{figure}[htbp]
\centerline{\includegraphics[width=0.6\textwidth]{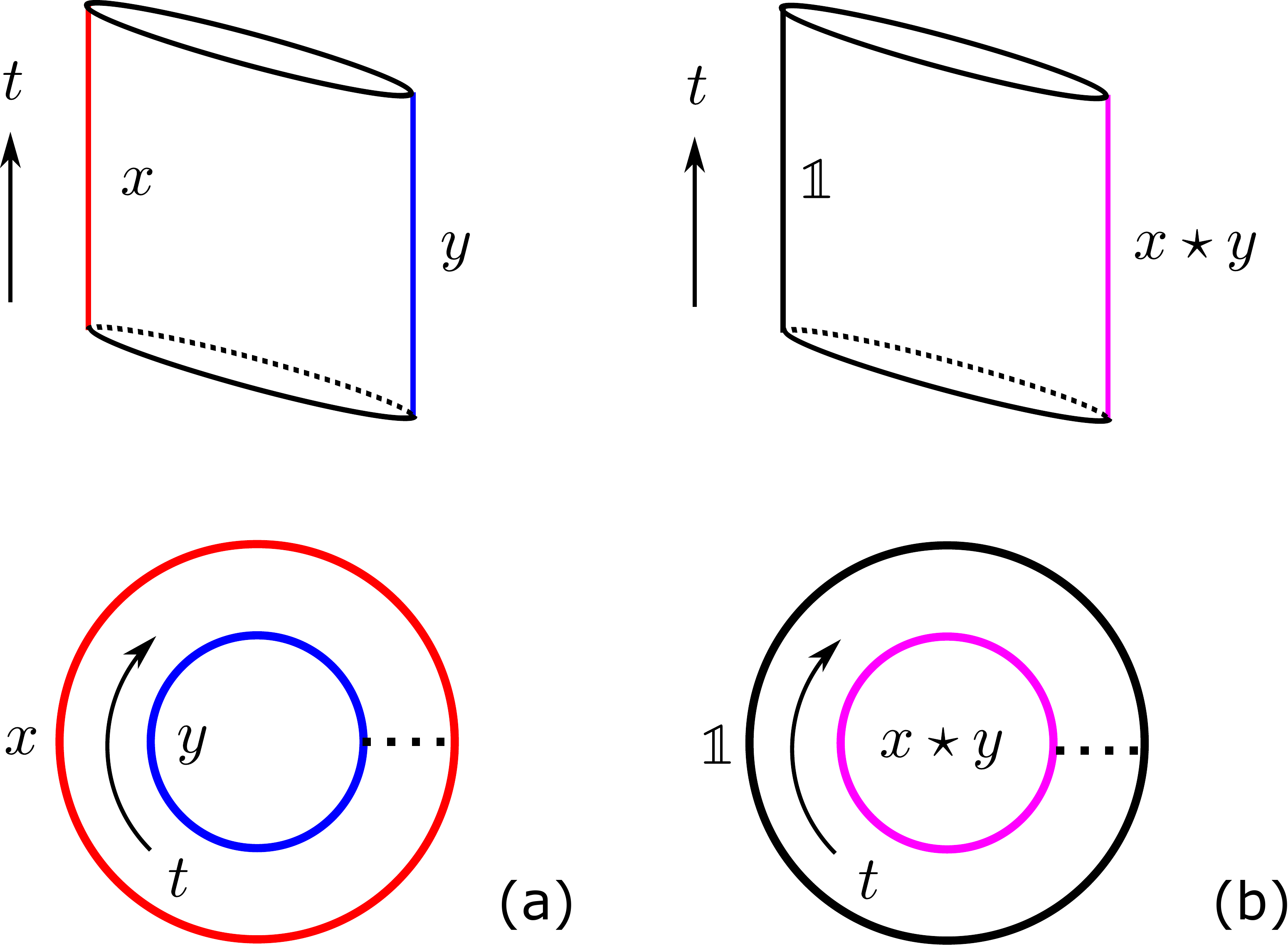}}
\caption[]{\label{fig:partition} Calculation of the partition function of $M_{x,y}$ via path integrals}
\end{figure}

\medskip
It is enough to just recover $V$ and the partition function of $M_{x,y}$. 
In physics, the partition function of $M_{x,y}$ can be computed by the path
integral over an annulus obtained by compactifying the time axis, and fixing
the two boundary conditions to be $x$ and $y$ as shown in Figure\, \ref{fig:partition}. The pictures in Figure\, \ref{fig:partition} also show that the partition function of $M_{x,y}$ is the same as that of $M_{\mathbb{1}, x\star y}$. This is also obvious by the fact that $M_{x,y}=M_{\mathbb{1}, x\star y}$ as objects in $Z(\Is)$. We denote the partition function of $M_{\mathbb{1}, x}=x$ by $Z_x(\tau)$ for $x=\mathbb{1}, e, m, f, \chi_\pm$, where $\tau$ is the moduli of torus.

Since all $x=\mathbb{1}, e, m, f, \chi_\pm$ are obtained from double Ising topological order via an anyon condensation, they can all be viewed as objects in $\Mod_V\boxtimes \overline{\Mod_V}$ as shown in Eq.\,(\ref{eq:1emf}) and (\ref{eq:chi}). Therefore, the partition function $Z_x(\tau)$ can be expressed in terms of the characters of $\one, \psi, \sigma$ in the Ising CFT, which are denoted by $\chi_0(\tau), \chi_{\frac{1}{2}}(\tau), \chi_{\frac{1}{16}}(\tau)$, respectively. Notice that $0,\frac{1}{2}, \frac{1}{16}$ are the lowest conformal weights of $\one, \psi, \sigma$, respectively. By Eq.\,(\ref{eq:1emf}) and (\ref{eq:chi}), we obtain six partition functions $Z_x(\tau)$ for $x=\mathbb{1}, e, m, f, \chi_\pm$: 
\begin{align}
&\chi_{\mathbb{1}}(\tau) = |\chi_0(\tau)|^2 + |\chi_{\frac{1}{2}}(\tau)|^2, \quad
\chi_e = \chi_m = |\chi_{\frac{1}{16}}(\tau)|^2, \quad
\chi_f = \chi_{\frac{1}{2}}(\tau)\chi_0(\tau)^\ast + \chi_0(\tau)\chi_{\frac{1}{2}}(\tau)^\ast, \\
&\hspace{1cm} 
\chi_{\chi_+}(\tau) = \chi_0(\tau)\chi_{\frac{1}{16}}(\tau)^\ast + \chi_{\frac{1}{2}}(\tau)\chi_{\frac{1}{16}}(\tau)^\ast, \quad\quad
\chi_{\chi_-}(\tau) = \chi_{\frac{1}{16}}(\tau)\chi_0(\tau)^\ast + \chi_{\frac{1}{16}}(\tau)\chi_{\frac{1}{2}}(\tau)^\ast. 
\end{align}

We will show in Section\,\ref{sec:lattice} explicitly how to recover this six partition functions $Z_x(\tau)$ (as summarized in Table\,\ref{tab:Z}). The partition functions of $M_{x,y}$ are automatically recovered for exactly the same reason (as illustrated in Figure\, \ref{fig:partition}), which is also manifest in the lattice model construction.

\section{A Lattice Model Realization} \label{sec:lattice}
In this section, we give a lattice model realization of the critical point of the pure edge phase transition between two different gapped edges of the 2d $\Zb_2$ topological order, and recover all the ingredients of the gappable gapless edge constructed in Section\,\ref{sec:edge}. We choose the Wen plaquette model to be the lattice realization of the 2d $\Zb_2$ topological order. For the convenience of readers in mathematical background, we review the Wen plaquette model in details.

\subsection{Wen Plaquette Model} \label{sec:wen-model}
\begin{figure}[htbp]
\begin{center}
\includegraphics[width=0.5\columnwidth]{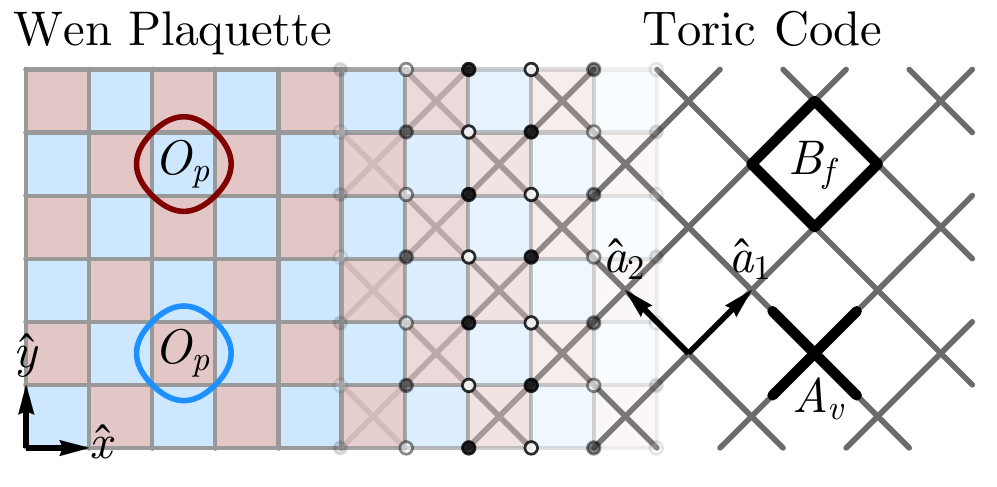}
\caption{Wen plaquette model on the square lattice (left) and its connection to the toric code model (right). The plaquette operator $O_p$ maps to either the vertex operator $A_v$ or the plaquette (face) operator $B_f$ via local basis transformations of $\sigma^x\leftrightarrow\sigma^z$ on the hollow sites.}
\label{Fig:WPLattice}
\end{center}
\end{figure}

The Wen plaquette model is defined on a square lattice with a two-dimensional local Hilbert space on each site, or one qubit per site. The Hamiltonian of this model is given by 
\begin{align}\label{eq:WPmodel}
H_{\rm wp} = -\sum_p O_p,
\end{align}
where $p$ labels the plaquettes on the square lattice, as shown in Fig. \ref{Fig:WPLattice} (a). The operator $O_p$ associated to the plaquette $p$ acts on the four sites located at the four corners of the plaquette as
\begin{align}\label{eq:Op}
O_p=\diag{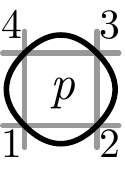}{34pt}=\sigma_1^z\sigma_2^x\sigma_3^z\sigma_4^x.
\end{align}
where $\sigma_i^x$ and $\sigma_i^z$ are the qubit operators (represented as Pauli matrices) and the subscript $i$ labels the site where these qubit operators are acting on. Here we have adopted the diagrammatic representation introduced in \cite{YW}, where each operator acting on a site is represented by
a string going through that site, i.e. $\sigma_i^z\equiv{\diag{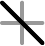}{10pt}}_i$
and $\sigma_i^x\equiv{\diag{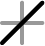}{10pt}}_i$. They anticommute with each other on the same site
$\sigma_i^x\sigma_i^z=-\sigma_i^z\sigma_i^x$, which can be diagrammatically represented as
$\diag{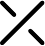}{10pt}=-\diag{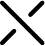}{10pt}$ (the operator that acts later will be stacked above). In Fig. \ref{Fig:WPLattice}, the lattice is partitioned into red and blue plaquettes, but the operators $O_p$ are defined identically for both types of plaquettes. This model $H_\text{wp}$ is exactly solvable because for any pairs of plaquettes $p$ and $p'$, the operators $O_p$ and $O_{p'}$ commute with each other, i.e.
\begin{align}
[O_p , O_{p'}] = 0.
\end{align}
Every state $\ket{\rm GS}$ in the ground state Hilbert space of this model should satisfy 
\begin{align}
O_p\ket{\rm GS} = \ket{\rm GS}, \,\,\, \forall p.
\end{align} 
So $O_p$ are also called the stabilizers that stabilize the ground state. In fact, every eigenstate of $H_{\rm wp}$ is a simultaneous eigenstate of $O_p$ for all plaquettes $p$. By definition in Eq.\eqref{eq:Op}, the eigenvalues of $O_p$ are $\pm 1$. Excited states of $H_{\rm wp}$ are labeled by a set of $O_p$ eigenvalues containing $-1$. If an excited state carries a $-1$ eigenvalue of $O_p$ for a given plaquette $p$, we say the excited state contains an excitation at the plaquette $p$. 

The Wen plaquette model is in fact equivalent to the standard toric code model. To see this equivalence, we first re-identify the sites in the original square lattice (spanned by $\hat{x},\hat{y}$) as the centers of the links in a new square lattice (spanned by $\hat{a}_1,\hat{a}_2$) which is $45^\circ$ degree tilted from the original one as in shown in Fig. \ref{Fig:WPLattice}. A red plaquette of the original lattice will be associated to the sites in the new square lattice while a blue plaquette of the original lattice will be associated to the plaquette (face) of the new square lattice. We can transform the Wen plaquette model on the original lattice to the standard toric code model on the new tilted lattice. To do this, we perform a change of the basis on all the hollow sites (shown in Fig. \ref{Fig:WPLattice}) such that the role of $\sigma^x$ and $\sigma^z$ interchanges on those sites, i.e. $\sigma^x \leftrightarrow \sigma^z$. After this change of basis, the plaquette terms $O_p$ on the blue plaquettes become the standard plaquette term $B_f=\prod_{l\in\partial f}\sigma_l^x$ of the toric code model on the orange lattice. In the new basis, the plaquette terms $O_p$ on the red plaquettes become the standard vertex term $A_v=\prod_{l\in\dd v}\sigma_l^z$ of the toric code mode. The Hamiltonian in Eq. \eqref{eq:WPmodel} then becomes $H=-\sum_{v}A_v-\sum_{f} B_f$ on the new lattice, which is the toric code model Hamiltonian.

\begin{figure}[htbp]
\begin{center}
\includegraphics[width=0.9\columnwidth]{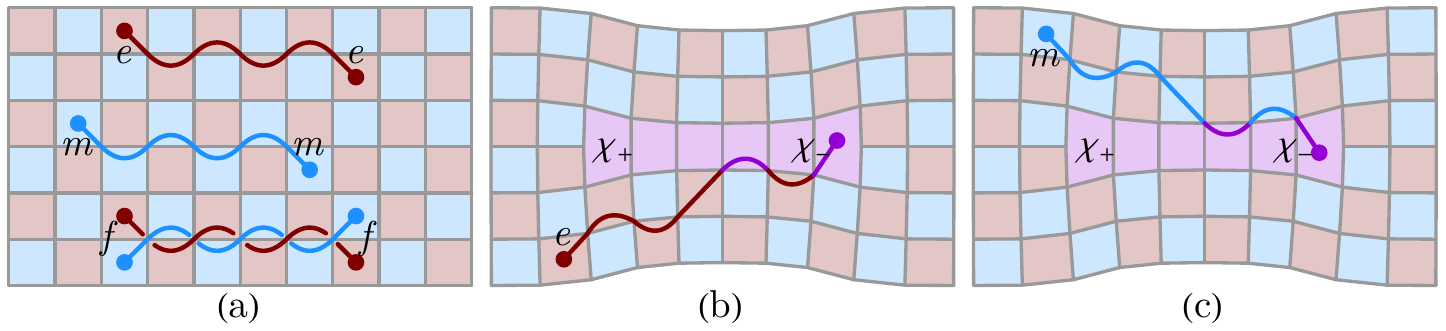}
\caption{(a) Intrinsic excitations are created in pairs by applying string operators to the ground state. Here the red string operator creates a pair of $e$ excitations, the blue string operator creates a pair of $m$ excitations, and the intertwined red and blue strings create a pair of $f$ excitations. (b,c) Extrinsic excitations involve modifying the lattice. Here we show a pair of dislocations: one plain dislocation $\chi_+$ and one excited dislocation $\chi_-$. $\chi_-$ can be obtained from $\chi_+$ adding an excitation of either $e$ or $m$ (they are indistinguishable in the presence of dislocation).}
\label{fig:strings}
\end{center}
\end{figure}

In the toric code model, the $A_v$ (or $B_f$) excitation is called the $e$ (or $m$) particle. Following this convention, in the Wen plaquette model, the red (or blue) plaquette excitation should be labeled as $e$ (or $m$) correspondingly. The bound state of $e$ and $m$ particles will be denoted as $f$, which is a fermion. These excitations can be created in pairs by string operators. The diagrammatic representation allows us to define string operators conveniently, simply by collecting the qubit operators along the string, see Fig.\,\ref{fig:strings}(a) for example. Together with the trivial particle $\mathbb{1}$ (representing a local excitation), $\mathbb{1},e,m,f$ form the set of simple objects in the toric code UMTC $\toric$.

Besides these intrinsic excitations, the Wen plaquette model also admits extrinsic excitations such as lattice dislocations \cite{bombin,kitaev-kong,YJW}. By an extrinsic excitation, we mean that the lattice dislocation is not an excited state in the spectrum of $H_{\rm wp}$ but rather a defect introduced to the system by modifying the Hamiltonian. The lattice dislocations are also created in pairs by removing a string of lattice sites, leaving two dislocation plaquettes at each end of the string, as shown in Fig.\,\ref{fig:strings}(b). The sites can be effectively removed by applying a strong external field to pin the qubits along the dislocation string, which amounts to adding the $H_\text{dis}=g\sum_{i\in\text{string}}\sigma_i^x$ term to the Wen plaquette model $H_\text{wp}$ \cite{YJW}. The plaquette operator around the dislocation will be extended to the following five-qubit operator
\begin{equation}\label{eq:Op5}
O_p=\ii\;\diag{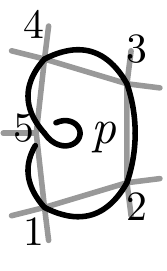}{48pt}=-\sigma_1^z\sigma_2^x\sigma_3^z\sigma_4^x\sigma_5^y,
\end{equation}
such that the local degeneracy around the dislocation can be lifted. The form of this dislocation plaquette operator can be derived from the $1/g$ perturbative expansion in the limit of strong pinning field $g\to\infty$ \cite{YJW}. The introduction of the dislocation does not break the exact solvability of the model as the extended plaquette operator $O_p$ (around the dislocation) still commute with all the rest of the plaquette operators. So all the eigenstate are still labeled by the eigenvalues of $O_p$ operators. In particular, around a dislocation plaquette $p$, $O_p=+1$ stabilizes a trivial dislocation denoted as $\chi_+$, while $O_p=-1$ indicates an excited dislocation denoted as $\chi_-$. Both dislocation defects $\chi_{\pm}$ implement the $e$-$m$ duality in the Wen plaquette model, as the lattice is distorted by the dislocation in such a way that there is no longer a global definition for red ($e$) and blue ($m$) plaquettes. If an $e$ particle goes around one dislocation $\chi_{\pm}$, it will transmute into an $m$ particle, and vice versa. Taking the dislocations $\chi_{\pm}$ into account, the six objects $\mathbb{1}, e,m,f,\chi_{\pm}$ are the simple objects in the UFC $Z(\Is)_A$. The following fusion rules are evident by drawing diagrams of strings/dislocations:
\begin{equation}
\begin{split}
& e\star e = m\star m = f\star f = \mathbb{1}, \quad e\star m = f, \\
&\chi_{\pm} \star \chi_{\pm} = \mathbb{1}\oplus f, \quad  \chi_{\pm} \star \chi_{\mp} = e\oplus m, \quad  
e\star \chi_{\pm} = \chi_{\pm} \star e = m\star \chi_{\pm} = \chi_{\pm} \star m = \chi_{\mp}.
\end{split}
\end{equation}
For example, Fig.\,\ref{fig:strings}(b,c) show how $\chi_+$ becomes $\chi_-$ by fusing with $e$ or $m$ (if we read the picture from left to right). They also illustrate $\chi_{\pm} \star \chi_{\mp} = e\oplus m$ (if we read the picture from top down or bottom up).

\subsection{Majorana Representation} \label{sec:majorana}

The Wen plaquette model also admits an alternative solution in terms of Majorana fermions. In this approach, the two-dimensional local Hilbert space $\Cb^2$ of a qubit on each site $i$ is first lifted to a four-dimensional super vector space $\Cb^{2|2}$, equipped with four Majorana fermion operators $\gamma_i^0,\gamma_i^1,\gamma_i^2,\gamma_i^3$. The Majorana operators are  Hermitian $\gamma_i^{a\dagger}=\gamma_i^a$ and satisfy the anticommutation relations $\{\gamma_i^a,\gamma_j^b\}=2\delta_{ij}\delta^{ab}$. We then use the on-site projections $\Cb^{2|2}\to\Cb^2$ specified by $\gamma_i^0\gamma_i^1\gamma_i^2\gamma_i^3=1$ to restrict the fermion Hilbert space to the physical (qubit) Hilbert space, which corresponds to the even-fermion-parity sector on each site. This construction amounts to first fractionalize the qubit into Majorana fermions and then impose the projective constraint to remove the gauge redundancy introduced in the fractionalization procedure.

Under the constraint $\gamma_i^0\gamma_i^1\gamma_i^2\gamma_i^3=1$, each qubit operator $\sigma_i^a$ ($a=x,y,z$) can be represented as Majorana fermion bilinear operators in two seemly different (but equivalent) ways:
\begin{equation}
\sigma_i^a=\ii\gamma_i^0\gamma_i^a=-\ii\epsilon^{abc}\gamma_i^b\gamma_i^c,
\end{equation}
where $a,b,c$ are used to label $x,y,z$ or $1,2,3$ interchangeably. It will be more intuitive to use the following diagrammatic representations
\begin{equation}
\arraycolsep=1.4pt\def\arraystretch{1}
\begin{array}{ccccc}
\diag{fig_sz}{12pt}&=&\diag{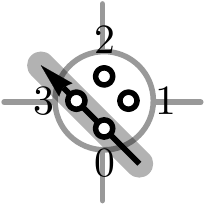}{36pt}&=&\diag{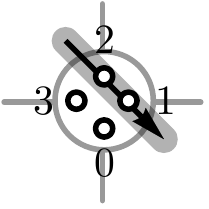}{36pt}\\
\rotatebox{90}{$\equiv$}& &\rotatebox{90}{$\equiv$}& &\rotatebox{90}{$\equiv$}\\
\sigma_i^z& &\ii\gamma_i^0\gamma_i^3& &\ii\gamma_i^2\gamma_i^1\end{array},\quad
\arraycolsep=1.4pt\def\arraystretch{1}
\begin{array}{ccccc}
\diag{fig_sx}{12pt}&=&\diag{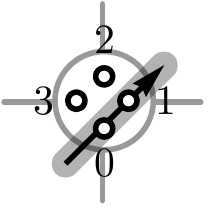}{36pt}&=&\diag{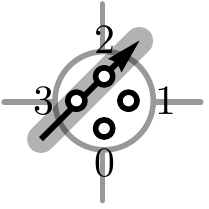}{36pt}\\
\rotatebox{90}{$\equiv$}& &\rotatebox{90}{$\equiv$}& &\rotatebox{90}{$\equiv$}\\
\sigma_i^x& &\ii\gamma_i^0\gamma_i^1& &\ii\gamma_i^3\gamma_i^2\end{array}.
\end{equation}
The four Majorana operators on each site are represented by small circles, and a string going through the site can pair up the fermion operators along the string in two different ways. The ordering of fermion operators is indicated by the arrow according to the following rules:
\begin{equation}
\diag[10pt]{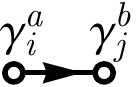}{16pt}\equiv\ii\gamma_i^a\gamma_j^b=-\ii\gamma_j^b\gamma_i^a\equiv-\,\diag[10pt]{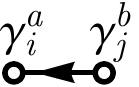}{16pt}.
\end{equation}
Using the Majorana representation, the plaquette operator $O_p$ can be written as
\begin{equation}
\begin{split}
O_p=\diag{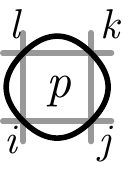}{34pt}&=\diag{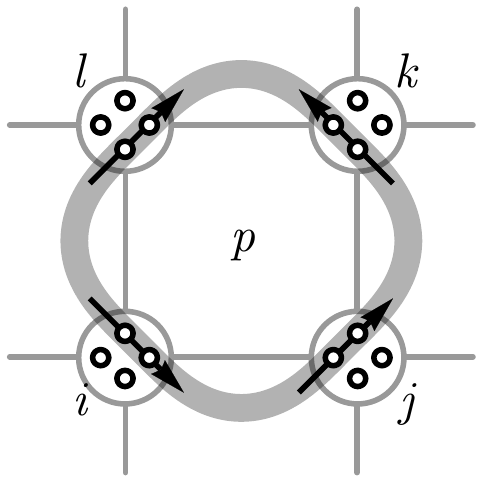}{80pt}\equiv(\ii\gamma_i^2\gamma_i^1)(\ii\gamma_j^3\gamma_j^2)(\ii\gamma_k^0\gamma_k^3)(\ii\gamma_l^0\gamma_l^1)\\
&=\diag{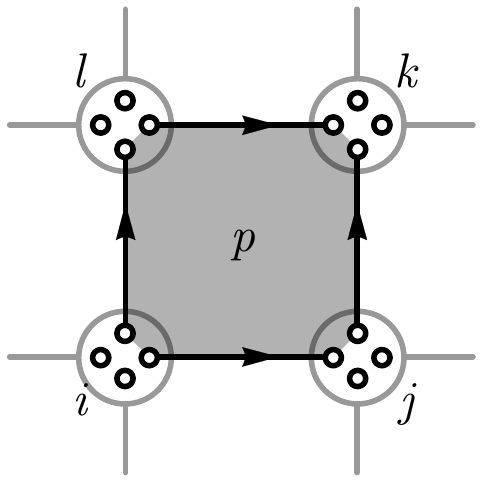}{80pt}\equiv(\ii\gamma_i^1\gamma_j^3)(\ii\gamma_j^2\gamma_k^0)(\ii\gamma_l^1\gamma_k^3)(\ii\gamma_i^2\gamma_l^0).
\end{split}
\end{equation}
It will be convenient to introduce the link operators (as Majorana fermion bilinear terms across each link)
\begin{equation}
\hat{\tau}_{i,i+\hat{x}}=\ii\gamma_i^1\gamma_{i+\hat{x}}^3,\quad \hat{\tau}_{i,i+\hat{y}}=\ii\gamma_i^2\gamma_{i+\hat{y}}^0,
\end{equation}
such that the plaquette operator $O_p=\prod_{\langle ij\rangle\in\partial p}\hat{\tau}_{ij}$ is simply a product of link operators around the plaquette. By definition, one can show that $\hat{\tau}_{ij}^\dagger=\hat{\tau}_{ij}$ and $\hat{\tau}_{ij}^2=1$, therefore $\hat{\tau}_{ij}$ only have two possible eigenvalues $\tau_{ij}=\pm1$. Moreover, the link operators $\hat{\tau}_{ij}$ commute with each other, so their eigenvalues $\tau_{ij}$ can be treated as independent $\Zb_2$ variables. In the common eigenbasis of $\hat{\tau}_{ij}$, the Wen plaquette model is diagonalized $H_\text{wp}=-\sum_{p}\prod_{\langle ij\rangle\in\partial p}\tau_{ij}$. If we identify the link variable $\tau_{ij}$ as a $\Zb_2$ gauge connection along $\langle ij\rangle$, $H_\text{wp}$ will describe a $\Zb_2$ gauge theory, which is invariant under the gauge transformation $\tau_{ij}\to s_i\tau_{ij}s_j$ (induced by any configuration of $s_i=\pm1$). The plaquette operator $O_p=\pm1$ measures the $\Zb_2$ gauge flux through each plaquette $p$ ($O_p=1$: no flux, $O_p=-1$: with flux). The Hamiltonian $H_\text{wp}=-\sum_pO_p$ energetically favors the $\Zb_2$ gauge flux to be trivial ($O_p=1$) everywhere for the ground state.

\begin{figure}[htbp]
\begin{center}
\includegraphics[width=\textwidth]{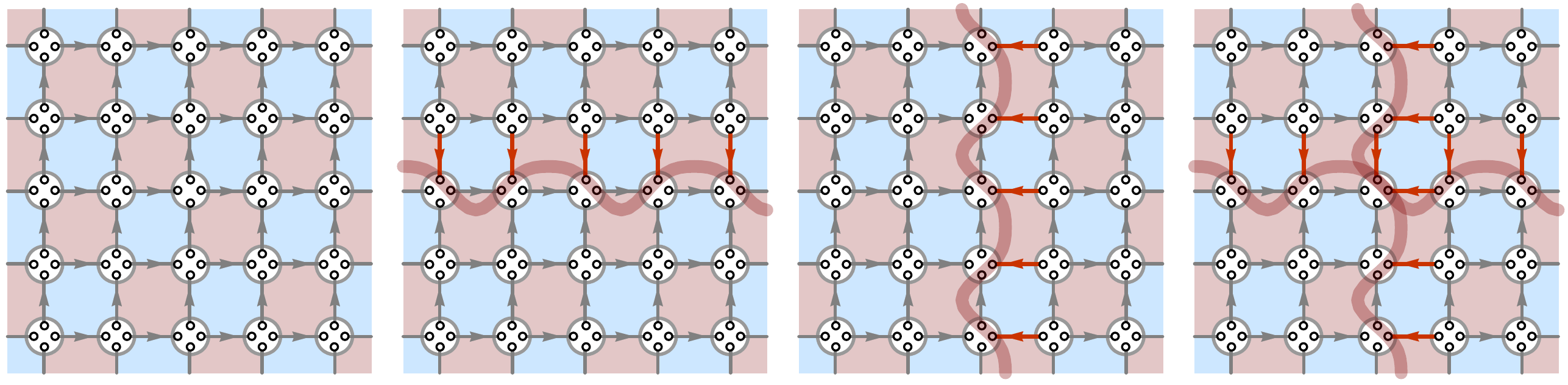}
\caption{Four gauge inequivalent configurations of $\tau_{ij}$ that satisfies $O_p=1$ for all plaquettes on a square lattice with periodic boundary condition in both directions. Link directions $i\to j$ to make $\tau_{ij}=+1$ are indicated by arrows, which point towards (against) $\hat{x}$ or $\hat{y}$ directions on the gray (red) links. These configurations are related to each other by applying string operators along non-contractable loops.}
\label{fig:configs}
\end{center}
\end{figure}

Assuming periodic boundary conditions of the square lattice along both $\hat{x}$ and $\hat{y}$ directions, there will be four gauge inequivalent configurations of $\tau_{ij}$ that solve the $O_p=1$ constraints (see Fig. \ref{fig:configs}), which correspond to the four degenerated ground states of the Wen plaquette model on a torus. Each configuration of $\tau_{ij}$ specifies a unique fermion state $\ket{\Psi[\tau_{ij}]}$ in the fermion Hilbert space, s.t. $\hat{\tau}_{ij}\ket{\Psi[\tau_{ij}]}=\tau_{ij}\ket{\Psi[\tau_{ij}]}$, where all Majorana fermions are dimerized in pairs and fully gapped. The physical ground state can be constructed by projection
\begin{equation}
\ket{\text{GS}}=\prod_{i}\frac{1+\gamma_i^0\gamma_i^1\gamma_i^2\gamma_i^3}{2}\ket{\Psi[\tau_{ij}]}=\sum_{[s_i]}\ket{\Psi[s_i\tau_{ij}s_j]}.
\end{equation}
Imposing the constraint $\gamma_i^0\gamma_i^1\gamma_i^2\gamma_i^3=1$ by projection is equivalent to summing over all gauge-related $\ket{\Psi[s_i\tau_{ij}s_j]}$ states (parameterized by $[s_i]$ that induces the gauge transform). Therefore the four gauge-inequivalent sectors of the $O_p=1$ subspace eventually result in the four physical ground states. The ground states are related to each other by string operators as shown in Fig.\ref{fig:configs}. In the Majorana representation, the string operator is proportional to the product of all Majorana operators covered by the string. Applying the string operator will change all $\gamma_i^a\to-\gamma_i^a$ along the string, and hence all $\tau_{ij}\to -\tau_{ij}$ if the link $\langle ij\rangle$ has an odd intersection with the string, thereby changing the state from one gauge sector to another.

\subsection{Edge Theory and Partition Functions}

\begin{figure}[htbp]
\begin{center}
\includegraphics[width=0.35\textwidth]{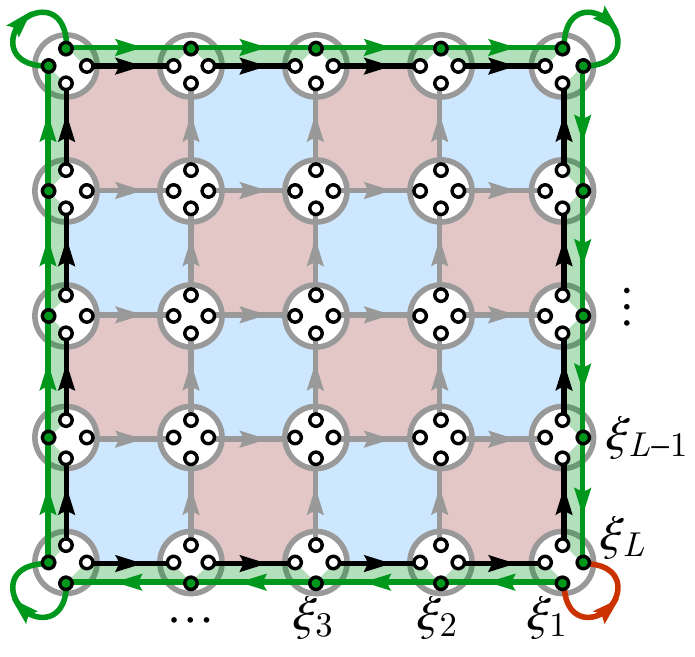}
\caption{The Majorana representation of the Wen plaquette model with open boundary. The dangling Majorana modes along the edge are marked as green dots. They are connected by the edge terms to form a Majorana chain coupled to the $\Zb_2$ gauge theory in the bulk. The red plaquettes are even ($e$) plaquettes and the blue plaquettes are odd ($m$) plaquettes.}
\label{fig:open}
\end{center}
\end{figure}

Now we consider placing the Wen plaquette model on a square lattice with open boundaries. The Majorana fermions in the bulk still pair up and remain gapped, however the dangling Majorana fermions along the edges of system can become gapless. In general, any term that commutes with the plaquette terms $O_p$ in the bulk can appear on the boundary, which will lift most of the edge degeneracy from the dangling Majorana modes and stabilize the edge theory. To the lowest order, the following edge terms will act on the links along the edge:
\begin{equation}\label{eq:edge_terms}
\begin{split}
\text{south edge: }&\sigma_i^z\sigma_j^x=\diag{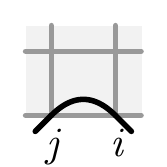}{32pt}=\diag{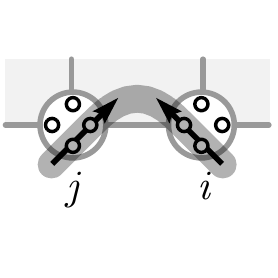}{60pt}=\diag{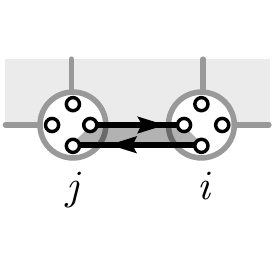}{60pt}=\ii\hat{\tau}_{ji}\gamma_i^0\gamma_j^0,\\
\text{west edge: }&\sigma_i^x\sigma_j^z=\diag{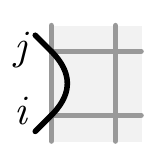}{32pt}=\diag{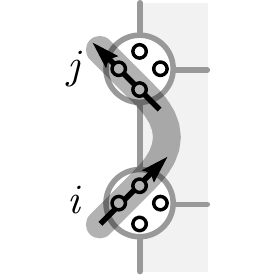}{60pt}=\diag{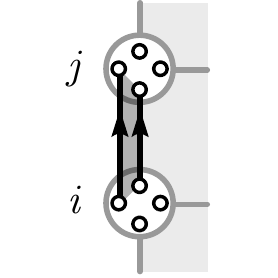}{60pt}=\ii\hat{\tau}_{ij}\gamma_i^3\gamma_j^3,\\
\text{north edge: }&\sigma_i^z\sigma_j^x=\diag{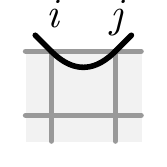}{32pt}=\diag{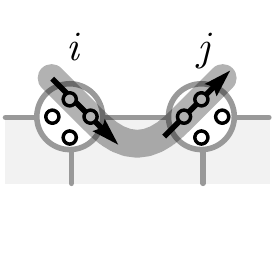}{60pt}=\diag{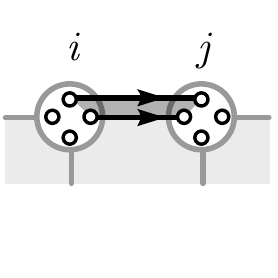}{60pt}=\ii\hat{\tau}_{ij}\gamma_i^2\gamma_j^2,\\
\text{east edge: }&\sigma_i^x\sigma_j^z=\diag{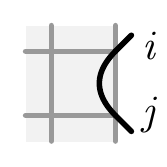}{32pt}=\diag{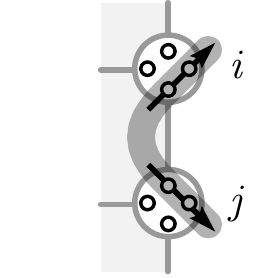}{60pt}=\diag{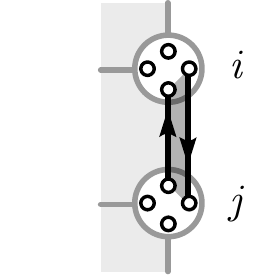}{60pt}=\ii\hat{\tau}_{ji}\gamma_i^1\gamma_j^1.
\end{split}
\end{equation}
They describe the dangling Majorana fermions hopping on the edge and coupling to the gauge connection $\hat{\tau}_{ij}=-\hat{\tau}_{ji}$ on the edge link. Moreover, the four corner sites of the square lattice allows the following corner terms:
\begin{equation}\label{eq:corner_terms}
\begin{split}
\text{south-west corner: }&\sigma_i^z=\diag{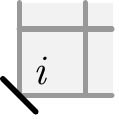}{24pt}=\diag{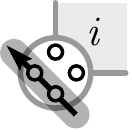}{28pt}=\diag{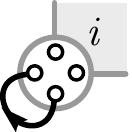}{28pt}=\ii\gamma_i^0\gamma_i^3,\\
\text{north-west corner: }&\sigma_i^x=\diag{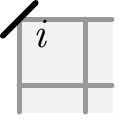}{24pt}=\diag{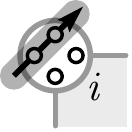}{28pt}=\diag{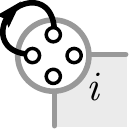}{28pt}=\ii\gamma_i^3\gamma_i^2,\\
\text{north-east corner: }&\sigma_i^z=\diag{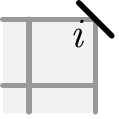}{24pt}=\diag{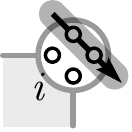}{28pt}=\diag{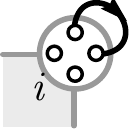}{28pt}=\ii\gamma_i^2\gamma_i^1,\\
\text{south-east corner: }&\sigma_i^x=\diag{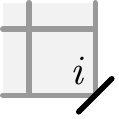}{24pt}=\diag{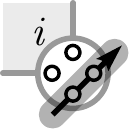}{28pt}=\diag{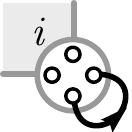}{28pt}=\ii\gamma_i^0\gamma_i^1.\\
\end{split}
\end{equation}
They describe how the Majorana fermion turns around at the corners. Gathering all these terms together and relabeling the dangling Majorana modes as $\xi_{i}$ (see Fig.\ref{fig:open}), the boundary of Wen plaquette model realizes a closed Majorana chain coupled to a $\Zb_2$ gauge field (originated from the bulk), which is described by the following edge Hamiltonian
\begin{equation}\label{eq:Hbdy}
H_\text{bdy}=-\sum_{i=1}^{L-1}\ii t_{i,i+1}\xi_i\xi_{i+1}-\ii t_{L,1}\xi_L\xi_1,
\end{equation}
where the total number of Majorana modes is $L=2(L_x+L_y)$ on a $L_x\times L_y$ lattice. The gauge connection $t_{ij}$ in $H_\text{bdy}$ is tied to the gauge connection $\tau_{ij}$ along the edge link according to Eq.\eqref{eq:edge_terms} as
\begin{equation}\label{eq:t=tau}
t_{ij}=\left\{\begin{array}{ll}
\tau_{ij}&\langle ij\rangle\text{ on the west and north edges},\\
\tau_{ji}=-\tau_{ij}&\langle ij\rangle\text{ on the east and south edges},\end{array}\right.
\end{equation}
For the corners, $t_{ij}$ are fixed by Eq.\eqref{eq:corner_terms} to be $t_{L1}=-1$ on the south-east corner and $t_{i,i+1}=+1$ on the rest of the corners (as illustrated in Fig.\ref{fig:open}). 

First of all, the edge is gapless. For any configuration of the gauge field $t_{ij}$, we can show that there always exist gauge invariant excitations in the spectrum of $H_\text{bdy}$ whose excitation energy $\Delta E$ lies below $4\pi/L$. Because along the one-dimensional edge, we can always gauge away the nontrivial gauge connection to the boundary condition, i.e. any configuration of $t_{ij}$ can be related to $t_{i,i+1}=1$ ($i=1,2,\cdots,L-1$) and $t_{L1}=\pm1$ by gauge transformation. Then Fourier transform to the momentum space $\xi_{k}=L^{-1/2}\sum_{i=1}^L \xi_i e^{-\ii k i}$, the boundary Hamiltonian in Eq.\eqref{eq:Hbdy} can be diagonalized as $H_\text{bdy}=\sum_{0<k<\pi}\epsilon_{k}\xi_{-k}\xi_{k}$, where $\epsilon_{k}=2\sin k$ describes the single-particle excitation energy $\epsilon_{k}$ as a function of the momentum $k$. Note that for Majorana fermions $\xi_{-k}=\xi_{k}^\dagger$, so $\xi_{k}$ at $k=0,\pi$ (if realizable) are Majorana operators (zero modes) and $\xi_{k}$ with $0<k<\pi$ are annihilation operator of fermion excitations. The excitation spectrum of $H_\text{bdy}$ can be labeled by fermion occupation numbers $n_k=\xi_{-k}\xi_{k}=0,1$ (for $0<k<\pi$) subject to the condition that the fermions can only be excited in pairs to maintain gauge invariance. Consider two-fermion excitations, the excitation energies are given by $\Delta E_{(k_1,k_2)}=\epsilon_{k_1}+\epsilon_{k_2}$ and labeled by two momenta $k_1$ and $k_2$. The boundary condition $t_{L1}=\pm1$ only affects the momentum quantization between $k=2\pi n/L$  (for $n=0,\cdots,L/2$) or $k=2\pi(n+1/2)/L$ (for $n=0,\cdots,L/2-1$). For $t_{L1}=+1$ (or $t_{L1}=-1$), we can explicitly construct the gauge invariant excitation of $\Delta E_{(0,2\pi/L)}=2\sin 2\pi/L$ (or $\Delta E_{(\pi/L,\pi/L)}=4 \sin \pi/L$). In either case, the minimal excitation energy is bounded by $\Delta E\leq4\pi/L$ at least. So the edge excitations gap vanishes (as $\sim1/L$) in the $L\to\infty$ thermodynamic limit.

Moreover, the edge is anomalous, in correspondence to the $\Zb_2$ topological order in the bulk. This is manifest from the fact that the edge Hamiltonian $H_\text{bdy}$ in Eq.\eqref{eq:Hbdy} is not a standalone one-dimensional model but has to involve the $\Zb_2$ gauge field from the bulk. To further expose the edge anomaly, we will study how the edge partition function responses to different bulk excitations (including both intrinsic and extrinsic excitations). This will in turn establish a bulk-edge correspondence between the  $\Zb_2$ topological order $Z(\Is)_A^0\simeq\toric$ in the bulk and its anomalous gapless edge $(V\otimes_\Cb\overline{V},Z(\Is),Z(\Is)_A)$ on the edge. 

We will start with intrinsic excitations $\mathbb{1},e,m,f$. To facilitate our discussion, we define the total $\Zb_2$ flux operator $(-)^\Phi$ and the edge fermion parity operator $(-)^F$ as
\begin{equation}\label{eq:PhiF}
(-)^\Phi=-t_{L1}\prod_{i=1}^{L-1}t_{i,i+1},\quad (-)^F=\prod_{r=1}^{L/2}\ii t_{2r-1,2r}\xi_{2r-1}\xi_{2r}.
\end{equation}
Both of them are invariant under gauge transformations $\xi_i\to s_i\xi_i$ and $t_{ij}\to s_it_{ij}s_j$ (for any $s_i=\pm1$). To make their physical meaning more explicit, we can use the gauge freedom to fix $t_{i,i+1}=1$ for $i=1,2,\cdots, L-1$ and push all the nontrivial gauge connection to the boundary condition $t_{L1}=\pm1$. After gauge fixing, $(-)^\Phi=-t_{L1}$ measures the total $\Zb_2$ gauge flux enclosed by the Majorana chain, such that $(-)^\Phi=1$ (or $(-)^\Phi=-1$) corresponds to the no flux (or $\pi$ flux) case, which also corresponds to the Neveu-Schwarz  (or Ramond) boundary condition for the Majorana fermion $\xi_i$. Also, with this gauge choice, $(-)^F=\prod_{r=1}^{L/2}\ii\xi_{2r-1}\xi_{2r}$ becomes the product of all dangling Majorana fermions $\xi_i$, which matches the definition of fermion parity in a purely one-dimensional Majorana chain. In conclusion, Eq.\eqref{eq:PhiF} provides a gauge-independent definition of the $\Zb_2$ flux $(-)^{\Phi}$ and edge fermion parity $(-)^F$ operators, which allow us to make connection to the plaquette operators in the bulk. Using the relation between $t_{ij}$ and $\tau_{ij}$ in Eq.\eqref{eq:t=tau} and using the fact that the total fermion parity of the system is even, it can be shown that
\begin{equation}\label{eq:PhiF=Op}
(-)^\Phi=\prod_{p}O_p,\quad (-)^F=\prod_{p\in\text{odd}}O_p.
\end{equation}
The even/odd plaquettes are such assigned that the plaquette on the south-east corner (the corner of $\xi_1$ and $\xi_L$) is always defined to be even, and the rest of the plaquettes can be labeled odd or even following the checkerboard pattern, see Fig.\ref{fig:open}. We will follow the convention to call the excitation in the even (odd) plaquette as the $e$ ($m$) excitation, and treat $f$ as the bound state of $e$ and $m$. Then according to Eq.\eqref{eq:PhiF=Op}, $(-)^\Phi$ counts the parity of $e$ and $m$ particles and $(-)^F$ counts the parity of $m$ and $f$ particles in the bulk. Therefore pushing the bulk particle to the edge can change the $\Zb_2$ flux and fermion parity of the Majorana chain, which is reflected in the change of the partition function of the edge theory. The results are summarized in Tab.\ref{tab:Z} and will be explained later in details. 

\begin{table}[htbp]
\caption{Relation between the excitations in the bulk and the partition function of the CFT on the edge.}
\begin{center}
\begin{tabular}{cccl}
 & $(-)^{\Phi}$ & $(-)^F$ & partition function\\
\hline
$\mathbb{1}$ & $+1$ & $+1$ & $Z_{\mathbb{1}}=|\chi_0(\tau)|^2 + |\chi_{\frac{1}{2}}(\tau)|^2$\\
$e$ & $-1$ & $+1$ & $Z_{e}=|\chi_{\frac{1}{16}}(\tau)|^2 $\\
$m$ & $-1$ & $-1$ &  $Z_{m}=|\chi_{\frac{1}{16}}(\tau)|^2 $\\
$f$ & $+1$ & $-1$ & $Z_{f}=\chi_0(\tau)^* \chi_{\frac{1}{2}}(\tau) + \chi_{\frac{1}{2}}(\tau)^* \chi_0(\tau)$\\
$\chi_+$ & $+1$ &  & $Z_{\chi_+}= \chi_{\frac{1}{16}}(\tau)^* \chi_{0}(\tau) +  \chi_{\frac{1}{16}}(\tau)^* \chi_{\frac{1}{2}}(\tau)$\\
$\chi_-$ & $-1$ &  & $Z_{\chi_-}= \chi_{0}(\tau)^* \chi_{\frac{1}{16}}(\tau)  +   \chi_{\frac{1}{2}}(\tau)^*\chi_{\frac{1}{16}}(\tau)$\\
\end{tabular}
\end{center}
\label{tab:Z}
\end{table}

To calculate these partition functions, we work with the $t_{i,i+1}=1$ gauge (for $i=1,2,\cdots,L-1$). The Majorana chain can be described by a free fermion CFT at low energy. As we discussed, the edge fermion dispersion relation is given by $\epsilon_{k}=2 \sin k$. Hence, the low energy fermions are described by the fermion modes with momenta $k=0+k_L$ and $k=\pi+k_R$ for small $k_L$ and $k_R$. As we will see, the fermion modes with small $k_L$ and $k_R$ correspond to the left-moving and right-moving fermion modes in terms of the 1+1D free fermion CFT. By linearizing the dispersion for small $k_L$ and $k_R$, we can write down the Hamiltonian $H$ and the total momentum $P$ in the momentum space:
\begin{equation}
\begin{split}
& H=\sum_{k_L}v_F k_L n_{k_L} + \sum_{k_R} -v_F k_R n_{k_R}, \\
& P=\sum_{k_L}k_L n_{k_L} + \sum_{k_R}k_R n_{k_R},
\end{split}
\label{eq:freefermionHP}
\end{equation}
where the Fermi velocity happens to be $v_F=2$ for the lattice model in Eq.\eqref{eq:Hbdy}. $n_{k_L}=0,1$ and $n_{k_R}=0,1$ denote the fermion occupation number of the momentum $k=0+k_L$ and $k=\pi+k_R$ modes. We notice that both the Hamiltonian $H$ and the total momentum $P$ receive contributions separately from the left- and right-moving fermion modes. 

Without a $\Zb_2$ flux in the bulk $(-)^{\Phi}=-t_{L1}=+1$, the fermion sees an antiperiodic boundary condition (as $t_{L1}=-1$) on the lattice and and its momentum is quantized to $k=2\pi(n+1/2)/L$ (for $n\in\Zb$). Since $L$ is even, the quantization of $k_L$ and $k_R$ are consequently given by $k_L = 2\pi(n'+1/2)/L$ and $k_R = 2\pi(n''+1/2)/L$ with $n',n''\in\mathbb{Z}$. This quantization of $k_L$ and $k_R$ implies that the left- and right-moving fermions are both subject to the Neveu-Schwarz (NS) boundary condition, namely the anti-periodic boundary condition in the CFT sense, in the spatial direction. Here, the boundary conditions for the left- and right-moving fermion modes are the same and are identical to the boundary condition defined on the lattice. However, one always needs to be cautious that these boundary conditions are not necessary equal. We will encounter such cases in the presence of the dislocations $\chi_{\pm}$.
 
The partition function on a spacetime torus with modular parameter $\tau=(\alpha+\ii\beta v_F)/L$ for both even and odd fermion parity will be given by 
\begin{equation}
\begin{split}
Z_{\mathbb{1}}(\tau)  &=\Tr_\text{NS} \frac{1+(-)^F}{2}e^{-\beta H+\ii \alpha P},\\
Z_{f}(\tau)&=\Tr_\text{NS} \frac{1-(-)^F}{2}e^{-\beta H+\ii \alpha P},
\end{split}
\end{equation}
where $\Tr_{\rm NS}$ represents the trace over the left- and right-moving fermion modes with the $k_L$ and $k_R$ quantization discussed above. These partition functions can be calculated by calculating the two terms $\Tr_\text{NS} e^{-\beta H+\ii \alpha P}$ and $\Tr_\text{NS} (-)^F e^{-\beta H+\ii \alpha P}$ separately. The form of the Hamiltonian $H$ and the total momentum $P$ given in Eq. \ref{eq:freefermionHP}, together with the fact that the total fermion parity operator $(-)^F$ is a product of the left and right fermion parities, ensures that each of $\Tr_\text{NS} e^{-\beta H+\ii \alpha P}$ and $\Tr_\text{NS} (-)^F e^{-\beta H+\ii \alpha P}$ factorizes into a product of the left-moving-fermion contribution and the right-moving-fermion contribution:
\begin{equation}
\begin{split}
& \Tr_\text{NS} ~e^{-\beta H+\ii \alpha P} = |d_{- -}(\tau)|^2, \\
& \Tr_\text{NS} ~(-)^F e^{-\beta H+\ii \alpha P} = |d_{- +}(\tau)|^2,
\end{split}
\end{equation}
where $d_{- \mp}(\tau)=q^{-\frac{1}{48}}\prod_{n=0}^{\infty}(1\pm q^{n+1/2})$ with $q=e^{2\pi\ii\tau}$ are the contributions from the left-moving fermions. The contributions from the right-moving fermions are given by $d_{- \mp}(\tau)^*$. Details of $d_{- \mp}(\tau)$ can be found in Chapter 6.4 and Chapter 10.3 in \cite{CFTbook}. The two subscripts of $d_{- \pm}(\tau)$ represents the spatial and temporal boundary conditions (in the CFT sense) respectively: ``$-$" represents the antiperiodic (Neveu-Schwarz) boundary condition and ``$+$" represents the periodic (Ramond) boundary condition. Here, the trace $\Tr_{\rm NS}$ has already set the spatial boundary condition for both of the left- and right-moving modes to be anti-periodic (Neveu-Schwarz). The temporal boundary condition is periodic (Ramond) if there is an fermion parity operator $(-)^F$ in the trace $\Tr_\text{NS}$. It is antiperiodic (Neveu-Schwarz) if without. Putting the results together, we have 
\begin{equation}
\begin{split}
Z_{\mathbb{1}}(\tau)  &=\Tr_\text{NS} \frac{1+(-)^F}{2}e^{-\beta H+\ii \alpha P}=\frac{|d_{--}(\tau)|^2+|d_{-+}(\tau)|^2}{2},\\
Z_{f}(\tau)&=\Tr_\text{NS} \frac{1-(-)^F}{2}e^{-\beta H+\ii \alpha P}=\frac{|d_{--}(\tau)|^2-|d_{-+}(\tau)|^2}{2}.
\end{split}
\end{equation}

With a $\Zb_2$ flux in the bulk $(-)^{\Phi}=-t_{L1}=-1$, the fermion sees an periodic  boundary condition (as $t_{L1}=+1$) on the lattice and its momentum is quantized to $k=2\pi n/L$ (for $n\in\Zb$). Given that $L$ is even, the momenta of left- and right-moving fermion modes are then correspondingly are quantized to $k_L=2\pi n'/L$ and $k_R=2\pi n''/L$ with $n',n''\in\mathbb{Z}$. This quantization of $k_L$ and $k_R$ implies that the left- and right-moving fermions are both subject to the Ramond (R) boundary condition, namely the periodic boundary condition in the CFT sense, in the spatial direction. Notice that this spatial periodic boundary condition allows for fermion zero modes at $k_L=0$ and at $k_R=0$. In this case, the partition function in both fermion parity sectors will be given by
\begin{equation}
\begin{split}
Z_{e}(\tau)&=\Tr_\text{R} \frac{1+(-)^F}{2}e^{-\beta H+\ii \alpha P}=\frac{|d_{+-}(\tau)|^2+|d_{++}(\tau)|^2}{2}, \\
Z_{m}(\tau)&=\Tr_\text{R} \frac{1-(-)^F}{2}e^{-\beta H+\ii \alpha P}=\frac{|d_{+-}(\tau)|^2-|d_{++}(\tau)|^2}{2},
\end{split}
\end{equation}
where $d_{+\mp}(\tau)=\frac{1}{\sqrt{2}}q^{\frac{1}{24}}\prod_{n=0}^{\infty}(1\pm q^n)$ with $q=e^{2\pi\ii\tau}$. Here, $\Tr_{\rm R}$ represents the trace over the left- and right-moving fermion modes with the $k_L$ and $k_R$ quantization given above. The $\frac{1}{\sqrt{2}}$ factor takes care of the state counting in the presence of the zero mode. Here, the two subscripts of $d_{+\mp}(\tau)$ again labels the spatial and temporal boundary conditions respectively. These results are obtained in a similar fashion as $Z_{\mathbb{1}}(\tau)$ and $Z_f(\tau)$. Details of $d_{+ \mp}(\tau)$ can also be found in Chapter 6.4 and Chapter 10.3 in \cite{CFTbook}. Interestingly, we notice that $d_{++}(\tau) =0 $ due to the $n=0$ contributions. Hence, 
\begin{equation}
Z_{e}(\tau) = Z_{m}(\tau).
\end{equation}

The characters $d_{\pm \pm}(\tau)$ are in fact the partitions of the 1+1D free fermions CFT on the torus with different boundary conditions (or equivalently different spin structures). They are related to the characters $\chi_0$, $\chi_{\frac{1}{16}}$, $\chi_{\frac{1}{2}}$ of the Ising CFT in the following way:
\begin{align}
\begin{split}
& \chi_0(\tau) = \frac{d_{--}(\tau) + d_{-+}(\tau)}{2}, \\
& \chi_{\frac{1}{16}}(\tau) = \frac{d_{+-}(\tau)}{\sqrt{2}}, \\
& \chi_{\frac{1}{2}}(\tau) = \frac{d_{--}(\tau) - d_{-+}(\tau)}{2}, 
\end{split}
\end{align}
Therefore, we can rewrite the partition functions $Z_{\mathbb{1}}$, $Z_{e}(\tau)$, $Z_{m}(\tau)$ and $Z_{f}(\tau)$ as
\begin{align}
\begin{split}
& Z_{\mathbb{1}} = |\chi_0(\tau)|^2 + |\chi_{\frac{1}{2}}(\tau)|^2, \\
& Z_{f} = \chi_0(\tau)^* \chi_{\frac{1}{2}}(\tau) + \chi_{\frac{1}{2}}(\tau)^* \chi_0(\tau), \\
& Z_{e}(\tau) = Z_{m}(\tau) = |\chi_{\frac{1}{16}}(\tau)|^2 . 
\end{split}
\end{align}

In fact, some more interpretation of the partition functions is in order. In defining $Z_{\mathbb{1}}$, $Z_{e}(\tau)$, $Z_{m}(\tau)$ and $Z_{f}(\tau)$, we only consider the low-energy edge degrees of freedom in the traces ``$\Tr_{\rm NS}$" and "$\Tr_{\rm R}$". However, these partition functions can still be interpreted as the partition functions for whole system (including the edge and the bulk degrees of freedom) with the bulk in the ``infinite-gap limit". The bulk energy gap is set by the energy scale of the bulk Hamiltonian $H_{\rm wp}$ while the edge energy scale is set independently by $H_{\rm bdy}$. For a finite-size system, if we take the bulk energy gap to be infinite without changing the edge energy scale, the whole system will be automatically in the sector with $(-)^\Phi = (-)^F=1$ at low energy. The partition function of the whole system then receives contributions only from the gapless edge states in the corresponding sector and becomes $Z_{\mathbb{1}}(\tau)$ in this limit. To obtain $Z_{m}(\tau)$ as the partition function of the whole system, we need to change the bulk Hamiltonian $H_{\rm wp}$ on a single odd plaquette $p$. By changing the sign of the coupling for the operator $O_p$ on the given (odd) plaquette $p$, we obtain a new Hamiltonian $H_{\rm wp}'$ whose infinite-gap limit automatically favors the sector with $(-)^\Phi=-1 $ and $(-)^F=-1$ at low energy. In this limit, the partition function of the whole system receives contributions only from the gapless edge states in the corresponding sector and becomes $Z_{m}(\tau)$. $Z_e(\tau)$ and $Z_f(\tau)$ can be obtained in a similar way as the partition functions of the whole system.

In the following, we discuss the behavior of these partition functions $Z_{\mathbb{1}}(\tau)$, $Z_f(\tau)$, $Z_{e}(\tau)$ and $Z_{m}(\tau)$ under modular transformations. 
The characters $d_{\pm \pm}(\tau)$ has simple modular transformation properties:
\begin{equation}
\mathcal{S}\text{-transformation}~ \tau \rightarrow -1/\tau  : \left\{ 
\begin{split}
& d_{+ +}(-1/\tau ) = d_{+ +}(\tau ) = 0, \\
& d_{+ -}(-1/\tau) = d_{- +}(\tau), \\
& d_{- +}(-1/\tau ) = d_{+ -}(\tau), \\
& d_{- -}(-1/\tau ) = d_{- -}(\tau), \\
\end{split}
\right.
\end{equation}
\begin{equation}
\mathcal{T}\text{-transformation}~ \tau \rightarrow \tau+1: \left\{ 
\begin{split}
& d_{+ +}(\tau+1) = d_{+ +}(\tau) = 0, \\
& d_{+ -}(\tau+1) = e^{\ii 2\pi/24} d_{+ -}(\tau), \\
& d_{- +}(\tau+1) = e^{-\ii 2\pi/48} d_{- -}(\tau), \\
& d_{- -}(\tau+1) = e^{-\ii 2\pi/48} d_{- +}(\tau), \\
\end{split}
\right.
\end{equation}
These transformations induce the modular transformations of $Z_{\mathbb{1}}(\tau)$, $Z_f(\tau)$, $Z_{e}(\tau)$ and $Z_{m}(\tau)$:
\begin{align}
\left(
\begin{array}{c}
Z_{\mathbb{1}}(-1/\tau) \\
Z_e(-1/\tau) \\
Z_m(-1/\tau) \\
Z_f(-1/\tau)
\end{array} 
 \right)
 = S 
 \left(
\begin{array}{c}
Z_{\mathbb{1}}(\tau) \\
Z_e(\tau) \\
Z_m(\tau) \\
Z_f(\tau)
\end{array} 
 \right), 
 ~~~~
 \left(
\begin{array}{c}
Z_{\mathbb{1}}(\tau+1) \\
Z_e(\tau+1) \\
Z_m(\tau+1) \\
Z_f(\tau+1)
\end{array} 
 \right)
 = T
 \left(
\begin{array}{c}
Z_{\mathbb{1}}(\tau) \\
Z_e(\tau) \\
Z_m(\tau) \\
Z_f(\tau)
\end{array} 
 \right),  
\label{eq:modularZ1emf}
\end{align}
where the matrices $S$ and $T$ are given by
\begin{align}
S = \frac{1}{2}
 \left(
\begin{array}{cccc}
1 & 1 & 1 & 1\\
1 & 1 & -1 & -1 \\
1 & -1 & 1 & -1 \\
1 & -1 & -1 & 1
\end{array} 
 \right), ~~~~~
T = 
 \left(
\begin{array}{cccc}
1 & 0 & 0 & 0\\
0 & 1 & 0 & 0 \\
0 & 0 & 1 & 0 \\
0 & 0 & 0 & -1
\end{array} 
 \right).
\end{align}
In fact, Eq.\,(\ref{eq:modularZ1emf}) does NOT uniquely determines $S$ and $T$. However, the $S$ and $T$ matrices given above matches exactly the modular $S$- and $T$-matrices of the $\Zb_2$ topological order. Another observation is that $Z_{\mathbb{1}} + Z_{m}$ (or $Z_{\mathbb{1}} + Z_{e}$) matches exactly the Ising CFT partition function on a torus which is modular invariant. This is not surprising because of Eq.\,(\ref{eq:1emf}) and (\ref{eq:1emf}).

\begin{figure}[htbp]
\begin{center}
\includegraphics[width=0.32\textwidth]{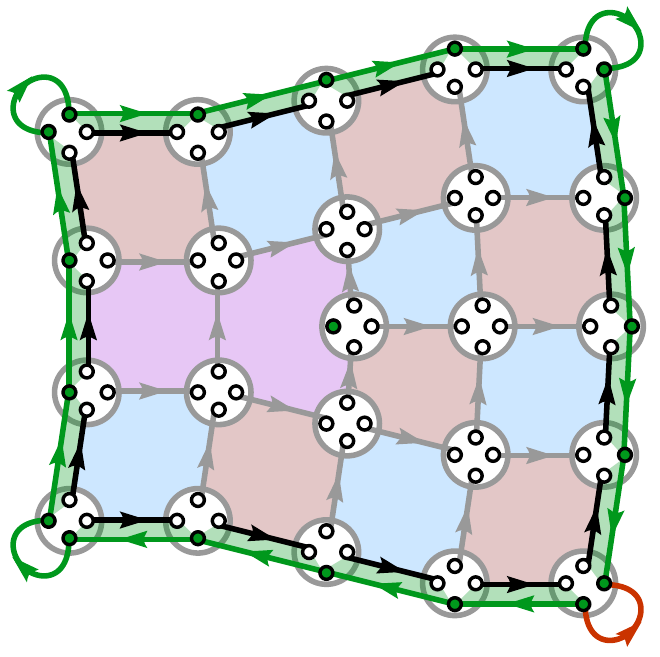}
\caption{The Majorana representation of the Wen plaquette model in the presence of a dislocation $\chi_+$. Green dots mark out the dangling Majorana modes along the edge as well as the Majorana zero mode trapped by the dislocation inside the bulk. The number of Majorana modes on the edge becomes odd. The even($e$)/odd($m$) plaquette can no longer be globally defined (the inconsistent plaquettes are colored in purple).}
\label{fig:dislocation}
\end{center}
\end{figure}

Now we turn to the extrinsic excitations $\chi_\pm$, which corresponds to lattice dislocations in the bulk. If we create a pair of lattice dislocations and push one of them through the edge, the number of lattice sites along the edge will be effectively reduced by one, resulting in an odd number of dangling fermions on the edge, as shown in Fig.\ref{fig:dislocation}. To compensate the missing Majorana mode on the edge, one Majorana zero mode will appear at the dislocation inside the bulk. In this case, the edge fermion parity operator can no longer be defined (as there are only odd number of fermion modes on the edge). This is also manifest from the bulk that the plaquettes can not be consistently assigned as even/odd in the presence of the dislocation. Therefore $(-)^F$ in Eq.\eqref{eq:PhiF=Op} no longer make sense. Nevertheless, the total $\Zb_2$ flux $(-)^\Phi$ is still well defined as $(-)^\Phi=-t_{L1}\prod_{i=1}^{L-1}t_{i,i+1}=\prod_pO_p$ as long as the plaquette operator is extended around the dislocation according to Eq.\eqref{eq:Op5}. Therefore, in the absence of excitations in the regular (non-dislocation) plaquettes, the trivial $\chi_+$ and excited $\chi_-$ dislocations in the bulk will lead to different boundary conditions for the boundary Majorana chain. Namely, $(-)^{\Phi}=\pm1$ corresponds to $\chi_\pm$. 

Interestingly, the presence of a single dislocation in the bulk renders the length $L$ of the edge Majorana chain odd. In this case, the boundary condition defined on the lattice for the edge Majorana chain is translated in the non-trivial way into the boundary condition (in the CFT sense) for the left- and right-moving fermion modes. Let's start with the dislocation $\chi_+$ which leads to $(-)^{\Phi}=+1$. The edge fermion modes experience an antiperiodic boundary condition on the lattice. Hence, their momenta are quantized to $k=2\pi(n+1/2)/L$ with $n\in\mathbb{Z}$. At low energy, we focus on the left- and right-moving fermion modes with momenta $k=0+k_L$ and $k_R=\pi + k_R$. When $L$ is odd, the quantization $k=2\pi(n+1/2)/L$ leads to the quantization $k_L = 2\pi(n'+1/2)/L $ and $k_R = 2\pi n''/L $ with $n',n''\in\mathbb{Z}$ for the left- and right-moving fermions. We notice that $k_L$ and $k_R$ are now quantized differently. These quantization suggests that the left-moving fermion modes are subject to an antiperiodic (Neveu-Schwarz) boundary condition in a CFT sense, while the right-moving fermions are subject to a periodic (Ramond) boundary condition in a CFT sense. Therefore, the partition function is given by
\begin{equation}
Z_{\chi_+}(\tau) = \frac{1}{\sqrt{2}} ~ d_{+-}(\tau)^* d_{--}(\tau),
\end{equation}
where $d_{--}(\tau)$ and $ d_{+-}(\tau)^*$ are the contributions from the left- and right-moving fermions. They are both subject to an antiperiodic boundary condition in the temporal direction. That is because there is no edge fermion parity projection needed and, hence, no edge fermion parity operator involved in obtaining $Z_{\chi_+}(\tau)$. $d_{--}(\tau)$ and $ d_{+-}(\tau)^*$ have the opposite spatial boundary conditions because the momentum quantization of $k_L$ and $k_R$ explained above. The reason that there is an extra factor of $\frac{1}{\sqrt{2}}$ in $Z_{\chi_+}(\tau)$ is a bit tricky. First of all, the edge Majorana chain with an odd length does not, strictly speaking, have a well-defined Hilbert space. One natural solution to it is to consider the total Hilbert space defined by both the edge fermion modes and the Majorana zero mode localized on the dislocation (even though the latter is decoupled from the former). This Majorana zero mode on the dislocation should contribute a $\sqrt{2}$ factor based on its quantum dimension. However, we need to ensure that the global fermion parity (including the edge and dislocation Majorana modes) is even (so that the states involved in the partition sum can be projected back to the ``bosonic" Hilbert space with just bosonic spin operators acting on it). The global fermion parity projection eliminates half of the states in the fermionic Hilbert space (associated to the Majorana modes) and hence leads to an extra factor of $1/2$. Therefore, the quantum dimension of the dislocation Majorana zero mode and the global fermion parity projection in total contribute to a factor of $\frac{\sqrt{2}}{2}$ to the partition function. Including the Majorana zero mode on the dislocation and implementing the global fermion parity projection are also natural if we consider $Z_{\chi_+}(\tau)$ as the partition function of the whole system (including the bulk and the edge) in the infinite-gap limit. That is because the Majorana zero mode localized on the dislocation has zero energy even when the bulk is in the infinite-gap limit. Furthermore, the global femion parity projection is always needed to ensure the states summed over in the partition function live in a bosonic Hilbert space.

We can analyze the case with a $\chi_-$ dislocation in a similar way. The dislocation $\chi_-$ which leads to $(-)^{\Phi}=-1$. Hence, the edge fermion modes experience a periodic boundary condition on the lattice. Their momenta are therefore quantized to $k=2\pi n/L$ with $n\in\mathbb{Z}$. At low energy, we focus on the left- and right-moving fermion modes with momenta $k=0+k_L$ and $k_R=\pi + k_R$. When $L$ is odd, the quantization $k=2\pi n /L$ leads to the quantization $k_L = 2\pi n'/L $ and $k_R = 2\pi (n''+1/2)/L $ with $n',n''\in\mathbb{Z}$ for the left- and right-moving fermions. Again, $k_L$ and $k_R$ are quantized differently. In the CFT sense, the left-moving fermion modes are now subject to a periodic (Ramond) boundary condition, while the right-moving fermions are subject to an antiperiodic (Neveu-Schwarz) boundary condition. Therefore, the partition function is given by
\begin{equation}
Z_{\chi_-}(\tau) = \frac{1}{\sqrt{2}} ~ d_{--}(\tau)^* d_{+-}(\tau).
\end{equation}
In terms of the Ising CFT characters, we can write 
\begin{align}
\begin{split}
& Z_{\chi_+}(\tau) =  \chi_{\frac{1}{16}}(\tau)^* \chi_{0}(\tau) +  \chi_{\frac{1}{16}}(\tau)^* \chi_{\frac{1}{2}}(\tau), \\
& Z_{\chi_-}(\tau) =  \chi_{0}(\tau)^* \chi_{\frac{1}{16}}(\tau)  +   \chi_{\frac{1}{2}}(\tau)^*\chi_{\frac{1}{16}}(\tau).
\end{split}
\end{align}
All the partition functions $Z_{\mathbb{1}}$, $Z_e$, $Z_m$, $Z_f$, $Z_{\chi_+}$, and $Z_{\chi_-}$ are summarized in Table \ref{tab:Z}. 

\medskip
We have recovered the partition functions of $M_{\mathbb{1}, x}$ for $x=\mathbb{1},e,m,f,\chi_{\pm}$ directly from the lattice model of the gapless edge. More general partition function of $M_{x,y}$ coincides with that of $M_{\mathbb{1}, x\otimes y}$. This is obvious from our lattice model construction. 
Therefore, the gapless edge defined by Eq.\,(\ref{eq:Hbdy}), which is coupled to the bulk, gives exactly the gapless edge $(V\otimes_\Cb\overline{V},Z(\Is),Z(\Is)_A)$. 

The partition functions $Z_{\mathbb{1}}$, $Z_e$, $Z_m$, $Z_f$, $Z_{\chi_+}$ obtained using a Hamiltonian formalism in this section are closely related to the results of \cite{amf} where the Ising CFT partition functions in the presence of topological line defects are studied using a 2-dimensional statistical mechanical model. The 2-dimensional statistical mechanical model proposed in \cite{amf} can be viewed as the discretized Euclidean path integral of the edge Hamiltonian studied in this section. The topological defect lines of the statistical mechanical can be identified as the worldline of the objects $x=\mathbb{1},e,m,f,\chi_{\pm}$ in our Hamiltonian formalism.

\subsection{Pure Edge Phase Transition}
In this subsection, we provide an interpretation of this gapless edge theory given by Eq.\,(\ref{eq:Hbdy}) as the critical point between two types of topologically distinct gapped edges of the $\Zb_2$ topological order. In the following, we only work with the Wen plaquette model with an open boundary and without any dislocations. The labeling of the edge Majonrana modes follows that of Fig. \ref{fig:open}. We use the same convention as above that the plaquette on the south-east corner is defined to be even. We consider a generalized edge Hamiltonian
\begin{equation}\label{eq:Hbdy'}
H_\text{bdy}'=-\sum_{i=1}^{L-1} (1+(-)^i \lambda) \ii t_{i,i+1}\xi_i\xi_{i+1}- (1+(-)^L \lambda) \ii t_{L,1}\xi_L\xi_1,
\end{equation}
where $\lambda$ is a tuning parameter. When $\lambda=0$, $H_\text{bdy}'$ reduces back to the Hamiltonian $H_\text{bdy}$ in Eq.\,(\ref{eq:Hbdy}) which gives rise to the gapless edge theory. A non-zero $\lambda$ results in an alternating pattern of the hopping strength in the edge Majorana chain. From now on, we define the links along the edge between the edge Majorana modes $\xi_{2r-1}$ and $\xi_{2r}$ with $r=1,2,...,L/2$ as the $e$-links, and the links between the edge Majorana modes $\xi_{2r}$ and $\xi_{2r+1}$ with $r=1,2,...,L/2-1$, as well as the link between $\xi_L$ and $\xi_1$, as the $m$-links. In the generalized edge model $H_\text{bdy}'$, the hopping strength along the $e$-links are given by $1-\lambda$, while the  hopping strength along the $m$-links are given by $1+\lambda$. One should always remember that each of the fermion hopping terms can be written in terms of the bosonic operators in Eq.\,(\ref{eq:edge_terms}) and Eq.\,(\ref{eq:corner_terms}). Therefore, $H_\text{bdy}'$ can be expressed using the Pauli matrices on the sites on edge.

For the simplicity of discussion, we can focus on the sector such that the edge Majorana chain has an antiperiodic boundary condition, i.e. $(-)^\Phi = 1$. In this case, we can gauge fix such that $t_{i,i+1}=t_{L,1}=1$ for all $i=1,2,...,L-1$. $H_\text{bdy}'$ then becomes the standard 1d Majorana chain. When $\lambda \neq 0$, the dispersion of the edge fermions is gapped and, hence, the Hamiltonian $H_\text{bdy}'$ is gapped as well. It is well-known that the phases with $\lambda>0$ and $\lambda<0$ are two distinct symmetry-protected-topological (SPT) phases of the 1d Majorana chain. The states with $\lambda>0$ (or $\lambda<0$) are all adiabatically connected. The two SPT phases each has a simple limit at $\lambda=1$ and at $\lambda=-1$ respectively. With $\lambda=1$, the hopping along the $e$-links are completely switched off and the edge Majorana modes dimerize on the $m$-links. With $\lambda=-1$, the hopping along the $m$-links are completely switched off and the edge Majorana modes dimerize on the $e$-links instead. One can easily see that the two SPT phases can be mapped into each other by exchanging the roles of the $e$-links and the $m$-links. Therefore, if we enforce a symmetry between $e$-links and the $m$-links in the edge model (which consequently requires $\lambda=0$ in $H_\text{bdy}'$), the edge theory has to be at the critical point between the two SPT phases. 

The two SPT phases on the edge and the critical point between them can be reinterpreted in connection to the $\Zb_2$ topological order in the bulk. For the phase with $\lambda>0$, we can take the case with $\lambda=1$ as a representative. When $\lambda=1$, the only terms left in $H_\text{bdy}'$ are the hopping terms on the $m$-links which are exactly the edge terms (shown in Eq.\,(\ref{eq:edge_terms})) that are adjacent to the odd plaquettes of the bulk.\footnote{In fact, some $m$-links are associated to the corner terms on the corner of even plaquettes. But it will not change the following discussion.} Remember that we have associated the $m$ particle to the odd plaquette in the bulk. Only turning on the edge terms adjacent to the odd plaquettes in fact enfores the $m$-condensing (gapped) edge of the toric topological order. Hence, the gapped edge phase with $\lambda>0$ should be associated to the $m$-condensing edge of the $\Zb_2$ topological order. Following a similar line of reasoning, we can conclude that the gapped  edge phase with $\lambda<0$ should be associated to the $e$-condensing edge of the $\Zb_2$ topological order. The gapless point at $\lambda=0$ is in fact the critical point between the two types of gapped edges of the $\Zb_2$ topological order. Generically, the edge of $\Zb_2$ topological order does not need to be at the critical point. However, if we enforce the $e$-$m$ duality as a ``symmetry" of the whole systems, there will be a ``symmetry" between the $e$-links and the $m$-links on the edge. It enforces a gapless edge that is the critical point between the two types of gapped edges.


\end{document}